\documentclass[iop, twocolumn]{aastex63}%linenumbers,
\usepackage{verbatim,subfigure}
\hypersetup{citecolor=blue,urlcolor=red}
\shorttitle{Pulsar timing}
\shortauthors{Zhao et al.}
\usepackage{ulem}
\usepackage[para]{threeparttable}
\begin{document}

\title{Follow-up timing of 12 pulsars discovered in Commensal Radio Astronomy FAST Survey}

\correspondingauthor{J.P. Yuan, N. Wang, D. Li}
\email{yuanjp@xao.ac.cn, na.wang@xao.ac.cn, dili@nao.cas.cn}

\author{D. Zhao}
\affiliation{Xinjiang Astronomical Observatory, Chinese Academy of Sciences, 150 Science 1-Street, Urumqi, Xinjiang 830011, China; \textcolor{blue}{na.wang@xao.ac.cn; yuanjp@xao.ac.cn; dili@nao.cas.cn}}
\affiliation{Key Laboratory of Radio Astronomy, Chinese Academy of Sciences, Urumqi, Xinjiang, 830011, China}
\affiliation{Xinjiang Key Laboratory of Radio Astrophysics, 150 Science1-Street, Urumqi, Xinjiang, 830011, China}

\author{J. P. Yuan\href{https://orcid.org/0000-0002-5381-6498}}%{\includegraphics[scale=0.04]}
\affiliation{Xinjiang Astronomical Observatory, Chinese Academy of Sciences, 150 Science 1-Street, Urumqi, Xinjiang 830011, China; \textcolor{blue}{na.wang@xao.ac.cn; yuanjp@xao.ac.cn; dili@nao.cas.cn}}
\affiliation{Key Laboratory of Radio Astronomy, Chinese Academy of Sciences, Urumqi, Xinjiang, 830011, China}
\affiliation{Xinjiang Key Laboratory of Radio Astrophysics, 150 Science1-Street, Urumqi, Xinjiang, 830011, China}

\author{N. Wang\href{https://orcid.org/0000-0002-9786-8548}}%{\includegraphics[scale=0.04]}} %\thanks{E-mail: na.wang@xao.ac.cn}
\affiliation{Xinjiang Astronomical Observatory, Chinese Academy of Sciences, 150 Science 1-Street, Urumqi, Xinjiang 830011, China; \textcolor{blue}{na.wang@xao.ac.cn; yuanjp@xao.ac.cn; dili@nao.cas.cn}}
\affiliation{Key Laboratory of Radio Astronomy, Chinese Academy of Sciences, Urumqi, Xinjiang, 830011, China}
\affiliation{Xinjiang Key Laboratory of Radio Astrophysics, 150 Science1-Street, Urumqi, Xinjiang, 830011, China}

\author{D. Li\href{https://orcid.org/0000-0003-3010-7661}}%{\includegraphics[scale=0.04]}
\affiliation{National Astronomical Observatories, Chinese Academy of Sciences, A20 Datun Road, Chaoyang District, Beijing 100101, China}
\affiliation{Xinjiang Astronomical Observatory, Chinese Academy of Sciences, 150 Science 1-Street, Urumqi, Xinjiang 830011, China; \textcolor{blue}{na.wang@xao.ac.cn; yuanjp@xao.ac.cn; dili@nao.cas.cn}}
\affiliation{University of Chinese Academy of Sciences, Beijing 100049, China}
\affiliation{Zhejiang Lab, Hangzhou, Zhejiang 311121, People’s Republic of China}

\author{P. Wang\href{https://orcid.org/0000-0002-3386-7159}}%{\includegraphics[scale=0.04]}}
\affiliation{National Astronomical Observatories, Chinese Academy of Sciences, A20 Datun Road, Chaoyang District, Beijing 100101, China}
\affiliation{Institute for Frontiers in Astronomy and Astrophysics, Beijing Normal University, Beijing 102206, China}

\author{M. Y. Xue\href{https://orcid.org/0000-0001-8018-1830}}%{\includegraphics[scale=0.04]}}
\affiliation{National Astronomical Observatories, Chinese Academy of Sciences, A20 Datun Road, Chaoyang District, Beijing 100101, China}

\author{W. W. Zhu\href{https://orcid.org/0000-0001-5105-4058}}%{\includegraphics[scale=0.04]}}
\affiliation{National Astronomical Observatories, Chinese Academy of Sciences, A20 Datun Road, Chaoyang District, Beijing 100101, China}
\affiliation{Institute for Frontiers in Astronomy and Astrophysics, Beijing Normal University, Beijing 102206, China}

\author{C. C. Miao\href{https://orcid.org/0000-0002-9441-2190}}%{\includegraphics[scale=0.04]}}
\affiliation{Zhejiang Lab, Hangzhou, Zhejiang 311121, People’s Republic of China}

\author{W. M. Yan\href{https://orcid.org/0000-0002-7662-3875}}%{\includegraphics[scale=0.04]}}
\affiliation{Xinjiang Astronomical Observatory, Chinese Academy of Sciences, 150 Science 1-Street, Urumqi, Xinjiang 830011, China; \textcolor{blue}{na.wang@xao.ac.cn; yuanjp@xao.ac.cn; dili@nao.cas.cn}}
\affiliation{Key Laboratory of Radio Astronomy, Chinese Academy of Sciences, Urumqi, Xinjiang, 830011, China}
\affiliation{Xinjiang Key Laboratory of Radio Astrophysics, 150 Science1-Street, Urumqi, Xinjiang, 830011, China}

\author{J. B. Wang\href{https://orcid.org/0000-0001-9782-1603}}%{\includegraphics[scale=0.04]}
\affiliation{Institute of Optoelectronic Technology, Lishui University, Lishui, 323000, China}
\affiliation{Xinjiang Astronomical Observatory, Chinese Academy of Sciences, 150 Science 1-Street, Urumqi, Xinjiang 830011, China; \textcolor{blue}{na.wang@xao.ac.cn; yuanjp@xao.ac.cn; dili@nao.cas.cn}}
\affiliation{Key Laboratory of Radio Astronomy, Chinese Academy of Sciences, Urumqi, Xinjiang, 830011, China}

\author{J. M. Yao\href{https://orcid.org/0000-0002-4997-045X}}%{\includegraphics[scale=0.04]}}
\affiliation{Xinjiang Astronomical Observatory, Chinese Academy of Sciences, 150 Science 1-Street, Urumqi, Xinjiang 830011, China; \textcolor{blue}{na.wang@xao.ac.cn; yuanjp@xao.ac.cn; dili@nao.cas.cn}}
\affiliation{Key Laboratory of Radio Astronomy, Chinese Academy of Sciences, Urumqi, Xinjiang, 830011, China}
\affiliation{Xinjiang Key Laboratory of Radio Astrophysics, 150 Science1-Street, Urumqi, Xinjiang, 830011, China}

\author{Q. D. Wu}
\affiliation{Xinjiang Astronomical Observatory, Chinese Academy of Sciences, 150 Science 1-Street, Urumqi, Xinjiang 830011, China; \textcolor{blue}{na.wang@xao.ac.cn; yuanjp@xao.ac.cn; dili@nao.cas.cn}}
\affiliation{Key Laboratory of Radio Astronomy, Chinese Academy of Sciences, Urumqi, Xinjiang, 830011, China}
\affiliation{Xinjiang Key Laboratory of Radio Astrophysics, 150 Science1-Street, Urumqi, Xinjiang, 830011, China}

\author{S. Q. Wang\href{https://orcid.org/0000-0003-4498-6070}}%{\includegraphics[scale=0.04]}}
\affiliation{Xinjiang Astronomical Observatory, Chinese Academy of Sciences, 150 Science 1-Street, Urumqi, Xinjiang 830011, China; \textcolor{blue}{na.wang@xao.ac.cn; yuanjp@xao.ac.cn; dili@nao.cas.cn}}
\affiliation{Key Laboratory of Radio Astronomy, Chinese Academy of Sciences, Urumqi, Xinjiang, 830011, China}
\affiliation{Xinjiang Key Laboratory of Radio Astrophysics, 150 Science1-Street, Urumqi, Xinjiang, 830011, China}

\author{S. N. Sun}
\affiliation{Xinjiang Astronomical Observatory, Chinese Academy of Sciences, 150 Science 1-Street, Urumqi, Xinjiang 830011, China; \textcolor{blue}{na.wang@xao.ac.cn; yuanjp@xao.ac.cn; dili@nao.cas.cn}}
\affiliation{Key Laboratory of Radio Astronomy, Chinese Academy of Sciences, Urumqi, Xinjiang, 830011, China}
\affiliation{Xinjiang Key Laboratory of Radio Astrophysics, 150 Science1-Street, Urumqi, Xinjiang, 830011, China}

\author{F. F. Kou\href{https://orcid.org/0000-0002-0069-831X}}%{\includegraphics[scale=0.04]}}
\affiliation{Xinjiang Astronomical Observatory, Chinese Academy of Sciences, 150 Science 1-Street, Urumqi, Xinjiang 830011, China; \textcolor{blue}{na.wang@xao.ac.cn; yuanjp@xao.ac.cn; dili@nao.cas.cn}}
\affiliation{Key Laboratory of Radio Astronomy, Chinese Academy of Sciences, Urumqi, Xinjiang, 830011, China}
\affiliation{Xinjiang Key Laboratory of Radio Astrophysics, 150 Science1-Street, Urumqi, Xinjiang, 830011, China}

\author{Y. T. Chen}
\affiliation{University of Chinese Academy of Sciences, Beijing 100049, China}
\affiliation{National Astronomical Observatories, Chinese Academy of Sciences, A20 Datun Road, Chaoyang District, Beijing 100101, China}

\author{S. J. Dang}
\affiliation{Guizhou Normal University, Guiyang 550001, China}

\author{Y. Feng\href{https://orcid.org/0000-0002-0475-7479}}%{\includegraphics[scale=0.04]}}
\affiliation{Zhejiang Lab, Hangzhou, Zhejiang 311121, People’s Republic of China}

\author{Z. J. Liu}
\affiliation{Guizhou Normal University, Guiyang 550001, China}

\author{X. L. Miao}
\affiliation{National Astronomical Observatories, Chinese Academy of Sciences, A20 Datun Road, Chaoyang District, Beijing 100101, China}

\author{L. Q. Meng\href{https://orcid.org/0000-0002-2885-568X}}%{\includegraphics[scale=0.04]}
\affiliation{National Astronomical Observatories, Chinese Academy of Sciences, A20 Datun Road, Chaoyang District, Beijing 100101, China}

\author{M. Yuan\href{https://orcid.org/0000-0003-1874-0800}}%{\includegraphics[scale=0.04]}}
\affiliation{National Astronomical Observatories, Chinese Academy of Sciences, A20 Datun Road, Chaoyang District, Beijing 100101, China}

\author{C. H. Niu\href{https://orcid.org/0000-0001-6651-7799}}%{\includegraphics[scale=0.04]}}
\affiliation{National Astronomical Observatories, Chinese Academy of Sciences, A20 Datun Road, Chaoyang District, Beijing 100101, China}
\affiliation{Institute of Astrophysics, Central China Normal University, Wuhan 430079, China}

\author{J. R. Niu\href{https://orcid.org/0000-0001-8065-4191}}%{\includegraphics[scale=0.04]}}
\affiliation{National Astronomical Observatories, Chinese Academy of Sciences, A20 Datun Road, Chaoyang District, Beijing 100101, China}

\author{L. Qian\href{https://orcid.org/0000-0003-0597-0957}}%{\includegraphics[scale=0.04]}}
\affiliation{National Astronomical Observatories, Chinese Academy of Sciences, A20 Datun Road, Chaoyang District, Beijing 100101, China}

\author{S. Wang\href{https://orcid.org/0000-0002-1570-7485}}%{\includegraphics[scale=0.04]}}
\affiliation{School of Computer Science, Fudan University, Shanghai 200438, China}

\author{X. Y. Xie}
\affiliation{Guizhou Normal University, Guiyang 550001, China}

\author{Y. F. Xiao}
\affiliation{GuiZhou University, Guizhou 550025, China}

\author{Y. L. Yue\href{https://orcid.org/0000-0003-4415-2148}}%{\includegraphics[scale=0.04]}}
\affiliation{National Astronomical Observatories, Chinese Academy of Sciences, A20 Datun Road, Chaoyang District, Beijing 100101, China}

\author{S. P. You}
\affiliation{Guizhou Normal University, Guiyang 550001, China}

\author{X. H. Yu}
\affiliation{Guizhou Normal University, Guiyang 550001, China}

\author{R. S. Zhao}
\affiliation{Guizhou Normal University, Guiyang 550001, China}

\author{R. Yuen}
\affiliation{Xinjiang Astronomical Observatory, Chinese Academy of Sciences, 150 Science 1-Street, Urumqi, Xinjiang 830011, China; \textcolor{blue}{na.wang@xao.ac.cn; yuanjp@xao.ac.cn; dili@nao.cas.cn}}
\affiliation{Key Laboratory of Radio Astronomy, Chinese Academy of Sciences, Urumqi, Xinjiang, 830011, China}
\affiliation{Xinjiang Key Laboratory of Radio Astrophysics, 150 Science1-Street, Urumqi, Xinjiang, 830011, China}

\author{X. Zhou\href{https://orcid.org/0000-0003-4686-5977}}%{\includegraphics[scale=0.04]}}
\affiliation{Xinjiang Astronomical Observatory, Chinese Academy of Sciences, 150 Science 1-Street, Urumqi, Xinjiang 830011, China; \textcolor{blue}{na.wang@xao.ac.cn; yuanjp@xao.ac.cn; dili@nao.cas.cn}}
\affiliation{Key Laboratory of Radio Astronomy, Chinese Academy of Sciences, Urumqi, Xinjiang, 830011, China}
\affiliation{Xinjiang Key Laboratory of Radio Astrophysics, 150 Science1-Street, Urumqi, Xinjiang, 830011, China}

\author{L. Zhang\href{https://orcid.org/0000-0001-8539-4237}}%{\includegraphics[scale=0.04]}}
\affiliation{National Astronomical Observatories, Chinese Academy of Sciences, A20 Datun Road, Chaoyang District, Beijing 100101, China}

\author{M. Xie}
\affiliation{Fudan University, Shanghai 200438, China}

\author{Y. X. Li}
\affiliation{Tencent Youtu Lab, 201103, Shanghai}

\author{C. J. Wang}
\affiliation{Tencent Youtu Lab, 201103, Shanghai}

\author{Z. K. Luo}
\affiliation{Tencent Youtu Lab, 201103, Shanghai}

\author{Z. Y. Gan}
\affiliation{Tencent Youtu Lab, 201103, Shanghai}

\author{Z. Y. Sun}
\affiliation{Tencent Youtu Lab, 201103, Shanghai}

\author{M. m. Chi}
\affiliation{Tencent Youtu Lab, 201103, Shanghai}

\author{C. J. Wang}
\affiliation{Tencent Youtu Lab, 201103, Shanghai}

\begin{abstract}

We present phase-connected timing ephemerides, polarization pulse profiles and Faraday rotation measurements of 12 pulsars
discovered by the Five-hundred-meter Aperture Spherical radio Telescope (FAST) in the Commensal Radio Astronomy FAST 
Survey (CRAFTS). The observational data for each pulsar span at least one year. Among them, PSR J1840+2843 shows subpulse drifting, and five pulsars are detected to exhibit pulse nulling phenomena. PSR J0640$-$0139 and PSR J2031$-$1254 are isolated MSPs with stable spin-down rates ($\dot{P}$) of $4.8981(6) \times $10$^{-20}$\,s\,s$^{-1}$ and $6.01(2) \times $10$^{-21}$\,s\,s$^{-1}$, respectively. Additionally, one pulsar (PSR J1602$-$0611) is in a neutron star - white dwarf binary system with 18.23-d orbit and a companion of  $\leq$  0.65M$_{\odot}$. PSR J1602$-$0611 has a spin period, companion mass, and orbital eccentricity that are consistent with the theoretical expectations for MSP - Helium white dwarf (He - WD) systems. Therefore, we believe it might be an MSP-He WD binary system. The locations of PSRs J1751$-$0542 and J1840+2843 on the  $P-\dot{P}$ diagram are beyond the traditional death line. This indicates that FAST has discovered some low $\dot{E}$ pulsars, contributing new samples for testing pulsar radiation theories. We estimated the distances of these 12 pulsars based on NE2001 and YMW16 electron density models, and our work enhances the dataset for investigating the electron density model of the Galaxy.

\end{abstract}

\keywords{surveys --  pulsars: general.}

\section{INTRODUCTION}

Radio pulsars are high-speed rotating neutron stars (NS) that are observed with periodic radio pulses. Because of its ultra-high density, strong magnetic field and ultrafast rotation frequency, 
it has become a natural laboratory for extreme physics.
Pulsars are used to study stellar evolution\citep{2017ApJ...846..170T} 
to constrain the equation of state of ultra-dense matter \citep{2001ApJ...550..426L,2016ARA&A..54..401O}, and to map  the distribution of free electrons in our Galaxy \citep{2002astro.ph..7156C,2017ApJ...835...29Y}.
The discovery of low spin-down energy loss rate ($\dot{E}$) pulsars provides the possibility to test radiation theory.
In the magnetic dipole frame,
the pulsar period ($P$) and period derivative $\dot{P}$ can be used to estimate the spin-down luminosity ($\dot{E}$), 
characteristic age ($\tau_c$), surface magnetic field ($B_s$) of pulsar. The death line of a pulsar defines the boundary at which the pulsar ceases radio emissions. When the pulsar is below the death line, it is not expected to produce radio emission any longer \citep{1975ApJ...196...51R,1993ApJ...402..264C,2000ApJ...531L.135Z}. By incorporating the general relativistic frame dragging \citep{1979ApJ...231..854A} and considering the curvature radiation and inverse Compton scattering, \citet{2000ApJ...531L.135Z} proposed the vacuum gap model and space-charge-limited-flow model. The detection of pulsars with low-$\dot{E}$ and long-period pulsars plays a crucial role in advancing our understanding of the radiation mechanisms inherent to pulsars.

%Radio radiation disappears when gaps cannot form above the polar cap area.

Radio millisecond pulsars (MSPs) are very stable rotators which 
are characterised by short spin periods ($P <$ 20
ms ) and very small spin-down rates ( $\dot{P} $ $\sim$ 10$^{-20}$\,s\,s$^{-1}$  ).
MSPs are believed to form in binary systems in which
the pulsars gain mass and angular momentum from the companion during the accretion process \citep{1991PhR...203....1B}. 
They are thus recycled and spun-up to milliseconds, and exhibit high rotational stability, manifested as accurate clocks in the Universe.
Their clock-like periodic pulses from binary pulsars can be used to determine their orbital dynamics with high precision. Excluding those MSPs in Globular clusters, which may be formed through dynamic processes, all MSPs should be in binary systems. However, approximately 20$\%$ of MSPs are isolated in the Galactic field. \citet{1988Natur.334..227V} and \cite{1988Natur.334..225K} proposed that the donor stars may have been ablated by the $\gamma$-rays and energetic particles emitted by the MSPs. \citet{2013ApJ...775...27C} suggests that the evaporation timescale may be too long unless a very high evaporation efficiency ($\sim$0.1).
\citet{2020A&A...633A..45J}  proposed a new process to form isolated Millisecond pulsar. When the central density of NS rises above the quark deconstrained density, a phase transition from NSs to strange stars may occur, and a suitable kick could disrupt the binary.
Depending on the presence or absence of a companion star, MSPs are divided into binary or isolated pulsars. For millisecond pulsars with stable rotation and very high timing accuracy, they can also constitute the Pulsar Timing Array (PTA). 
The timing array can not only be used to study the basic physics of neutron star,
but also be a unique tool to detect the ultra-low frequency gravitational waves generated by galaxy mergers and
study the basic characteristics of ultra-low frequency gravitational waves~\citep{2023ApJ...951L..10A,2023arXiv230713797A,2023ApJ...956L...3A,2023ApJ...951L...8A,2023ApJ...951L...9A,2023ApJ...951L..50A,2023ApJ...951L...6R,2023RAA....23g5024X,2023A&A...678A..48E,2023A&A...678A..49E,2023A&A...678A..50E}.
MSPs and binary pulsars provided an opportunity for studying the evolution of stellar and binary~\citep{1991PhR...203....1B,2022ASSL..465..201D,2023hxga.book..129B,2024arXiv240104395W}.

Pulsars exhibit highly polarized radio emissions~\citep{1968Natur.218..124L}, and studying their polarization properties helps revealing the radiation mechanisms and the geometry of the emission regions. Their high linear polarization allows us to measure Faraday rotation (RM), which aids in investigating the structure of the Galactic magnetic field~\citep{1974ApJ...188..637M,2006ApJ...642..868H}. Observations show that many pulsars exhibit an S-shaped variation in polarization position angle, which can be described by the Rotating Vector Model (RVM)~\citep{1969ApL.....3..225R}, allowing us to derive the pulsar’s geometric parameters~\citep{2023RAA....23j4002W}. Circular polarization is usually weak and exhibits sense reversal when the line of sight passes through the core component of the profile~\citep{1983ApJ...274..333R}. Additionally, some pulsars exhibit discontinuous changes in position angle, sometimes with 90-degree jumps, known as orthogonal polarization modes (OPM)~\citep{1975ApJ...196...83M,1984ApJS...55..247S,2022A&A...657A..34P}. In this work, we measured the polarization information of 12 CRAFTS pulsars. The high sensitivity of FAST will facilitate a more detailed study of pulsar polarization profiles.

Pulse nulling is a unique phenomenon in pulsar emissions, characterized by the sudden disappearance of the pulse signal for a period of time (e.g., \citealt{2007MNRAS.377.1383W}; \citealt{2020A&A...644A..73W}). Although the phenomenon of pulse nulling has been detected for many years, its emission physics is still unclear.
Several models have been proposed to explain the nulling phenomenon.
\citet{2010Sci...329..408L} proposed that variations in relativistic plasma flow are the underlying reason for a absence of emission during a null.
\citet{2010MNRAS.408..407B} proposed a partial shielding gap model. When the temperature of the polar cap reaches several thousand K,
the pulsar cannot emit pulse signals, showing the phenomenon of pulse nulling.
\citet{2007MNRAS.380..430H} suggested the periodic in nulling could be caused by a partially ignited sub-beam carousel.
\citet{2022MNRAS.517.1189O} propose that nulling may be due to the evolution of pair formation geometries producing core/cone emission and other emission effects.
At present, it is not clear how the variations of pulsar surface magnetosphere affect the modulation of pulsar emission intensity. Through long-term observations, it is also possible to study the causes of nulling phenomena and determine if there is periodic behavior.

The world's largest single-dish telescope, FAST \citep{2006ScChG..49..129N,2011IJMPD..20..989N,2020Innov...100053Q}, has discovered over 800 pulsars through the CRAFTS\footnote{\url{http://crafts.bao.ac.cn}} \citep{2018IMMag..19..112L}, GPPS\footnote{\url{http://zmtt.bao.ac.cn/GPPS/}} \citep{2021RAA....21..107H} and FAST Globular Cluster Pulsar Survey\footnote{\url{https://fast.bao.ac.cn/cms/article/65/}} \citep{2021ApJ...915L..28P}. It is expected that these surveys will discover more MSPs in the future, making a significant contribution to this population. Among them, CRAFTS discovered over 180 pulsars\footnote{\url{doi.org/10.57760/sciencedb.Fastro.00003}}, including 40 millisecond pulsars \citep{2021SCPMA..6429562W,2023MNRAS.522.5152W,2023MNRAS.518.1672M}, over 20 binary star systems \citep{2021MNRAS.508..300C,2023MNRAS.518.1672M}, 1 binary neutron star systems candidate~\citep{2023ApJ...958L..17W} and 1 confirmed binary neutron star system~\citep{2024ApJ...964L...7Z}, and one high energy young pulsar with $\gamma-$ray counterpart \citep{2021SCPMA..6429562W}.
\cite{2020MNRAS.495.3515C} reported the timing analysis of 11 isolated pulsars found in CRAFTS. The work lays the foundation for future efforts of CRAFTS.
%These pulsars were discovered by FAST during the commissioning phase using drift-scan mode and an ultra-wide bandwidth receiver, and confirmed and follow-up timed using the 64-m Parkes Radio Telescope.
\cite{2021MNRAS.508..300C} reported the timing analysis results of 10 pulsars found in the commissioning phase of FAST, 
one of which is a binary star.
This work indicates that the pulsar survey could enrich the pulsar population and improve the electron density model of the Galaxy.
\cite{2023MNRAS.518.1672M} reported 12 millisecond pulsars (MSPs) discovered with the FAST telescope in the CRAFTS. Eleven out of the twelve pulsars are in neutron star - Helium white dwarf binary systems, having companion masses consistent with the prediction of~\citet{1999A&A...350..928T} for MSPs.~\citet{2023MNRAS.522.5152W} reported 24 pulsars discovered with the FAST in CRAFTS and suggest that electron density models need updates for higher Galactic latitude regions.
Compared to pulsars discovered by other telescopes, CRAFTS can detect fainter and farther pulsars.

In this work, we report the timing results of 12 pulsars in FAST's CRFATS survey, 
including 9 isolated normal pulsars and three MSPs (two isolated MSPs PSRs J0640$-$0139 and J2031$-$1254 and one binary PSR J1602$-$0611). In Section~\ref{sec:obs}, we describe our observation setting and data reduction procedures.
We provide the results of follow-up timing observations in Section~\ref{sec:results}.
Our summary of the results is presented in Section~\ref{sec:conclusion}.

\section{OBSERVATIONS AND DATA PROCESSING}

\label{sec:obs} % used for referring to this section from elsewhere
These 12 pulsars were discovered by the NAOC-Tencent collaboration through the CRAFTS program \citep{2018IMMag..19..112L,2022arXiv220305738X} and are regularly monitored by FAST. The 19-beam receiver has an effective bandwidth of 400 MHz from 1050 to 1450 MHz \citep{2020RAA....20...64J}. The sampling rate of the observation is 49.152 $\mu$s. The observation data are recorded with full Stokes in the pulsar search mode. Each pulsar observation starts with a 40 seconds calibration noise diode for polarization calibration, with dish pointing at a sky region 10 arc-min offset from the source. Therefore, after deducting the time needed for antenna slew, the integration time for each pulsar is 240 seconds per epoch.

Unlike known pulsar, these newly discovered pulsars do not have ephemerides available to predict the arrival time of pulses.
Therefore, we first use the pulsar search analysis software PRESTO\footnote{\url{https://github.com/scottransom/presto}} \citep{2011ascl.soft07017R} to confirm the initial spin period (P) and dispersion measurements (DM) of the new pulsars.
For pulsars in binary systems, the apparent spin period and spin period derivative will significantly vary between epochs.
From the initial few months of follow-up observations,
we confirm that only one pulsar is in a binary system, and the other pulsars are isolated normal pulsars and MSPs.
For the pulsars in binary systems, the initial ephemerides needs to not only contains coordinates, estimated spin 
frequency ($\nu$), estimated spin frequency derivative ($\dot{\nu}$),
DM, but also consider Keplerian parameters describing the orbital motion.
We employ Java script newfitorbit\footnote{\url{https://github.com/SixByNine/newfitorbit}} to fit the Keplerian binary parameters: orbital period ($P_\mathrm{b}$), 
the epoch of periastron passage ($T_{0}$), and projected semi-major axis ($x$). In this process, we assumed that both the longitude of periastron ($\omega$) and the orbital eccentricity ($e$) are zero, resulting in a deviation from the initial ephemerides.
For the binary system in near circular orbit, it can be well described using the pulsar binary model ELL1 \citep{2001MNRAS.326..274L}.
In this model, the use of the time of ascending node ($T_\mathrm{asc}$) and Laplace-Lagrange parameters ($ \epsilon_{1} \equiv e$ $ \mathrm{sin}$ $\omega$, $ \epsilon_{2} \equiv e$ $ \mathrm{cos}$ $\omega$) in place of $T_{0}$, $\omega$ and $e$ is beneficial because employing $T_\mathrm{asc}$, $ \epsilon_{1}$ and $\epsilon_{2}$ helps to break the high correlation between  $T_{0}$ and $\omega$ in binary systems with small-eccentricity. The substitution makes it easier to obtain phase-connected solutions. Then,  according to the initial spin period and dispersion measure of the pulsar obtained, the pulses are folded with DSPSR\footnote{\url{http://dspsr.sourceforge.net/}} software \citep{2011PASA...28....1V}.
Integrated profiles with 512 phase bins are obtained for each epoch.
For the frequency channels and time sub-integrations by affected strong narrow-band and broad-band radio frequency interference (RFI), these channels and sub-integrations are zero-weighted with PSRCHIVE\footnote{\url{http://psrchive.sourceforge.net/}} software package \citep{2004PASA...21..302H}.
After aligning each observation and adding all observations,  we employ the PSRCHIVE routine \textbf{paas} to generate a standard profile and employ PSRCHIVE's \textbf{pat} tool to create times of arrival (TOAs).

\textbf{TEMPO} \citep{2015ascl.soft09002N} and \textbf{TEMPO2} \citep{2006MNRAS.369..655H} were applied to build a phase-connected
timing solution. 
We used the tool newfitorbit to obtain the initial timing solution of PSR J1602$-$0611, which is used as the initial ephemerides of DRACULA\footnote{\url{https://github.com/pfreire163/Dracula}} \citep{2018MNRAS.476.4794F} iteration, ultimately resulting in the accurate timing solution for PSR J1602$-$0611.
For the isolated pulsar, we fit the pulsar's positions, $\nu$, $\dot{\nu}$.
For the pulsar in binary systems, we fit the pulsar's positions, $\nu$, $\dot{\nu}$ and five Keplerian orbital parameters ($P_\mathrm{b}$, 
$T_\mathrm{asc}$, $x$, $ \epsilon_{1} $, $\epsilon_{2}$). 

We used the noise injection of each observation to calibrate the polarization,
and then select the observation with the highest signal-to-noise of each pulsar to search for RM using the rmfit tool, including the contribution of the Earth's ionosphere.
For each pulsar, we search in the range of $-$10000$\sim$10000 rad $\mathrm{m}^{-2}$, with a step size of 1. Furthermore, the DM was determined by dividing the entire frequency band into two sub-bands, each with a respective central frequency but the same bandwidth. The TOAs for these sub-bands were then fitted accordingly.
\begin{figure*}
	\includegraphics[width=7in,height=5in]{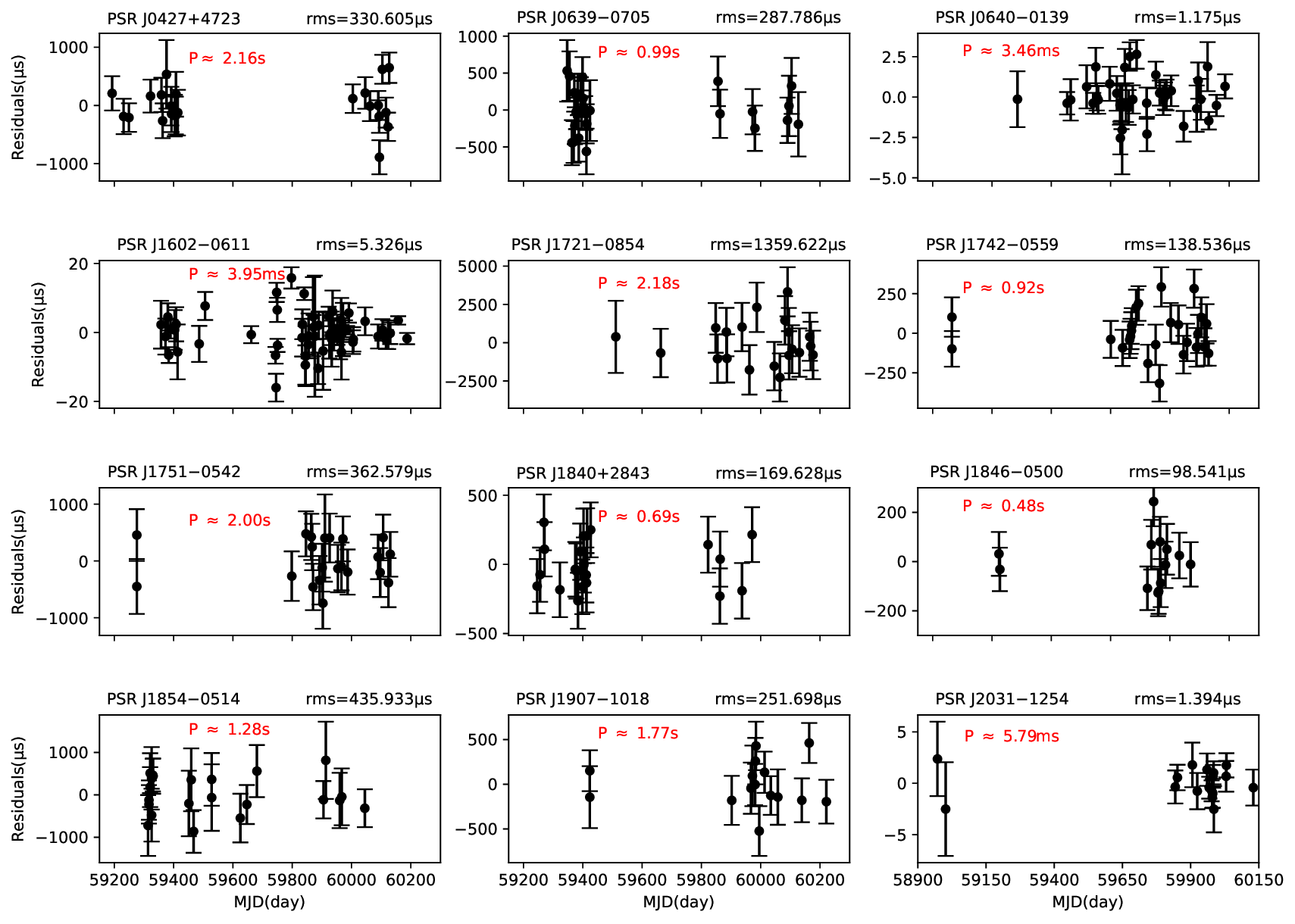}
    \caption{The best post-fit timing residuals of twelve pulsars. PSRs J0640$-$0139, J2031$-$1254 are isolated millisecond pulsars, and PSR J1602$-$0611 is a binary system. In the figure, the MJD range shown is 58900-60300 for PSR J2031$-$1254, while the MJD range for the other pulsars is 59150-60300.}
    \label{timing}
\end{figure*}
\subsection{Flux density}
\label{sec:flux} % used for referring to this section from elsewhere
We calculated the flux density of pulsar according to the following radiometer equation \citep{2004hpa..book.....L}:
\begin{equation}
    S_\mathrm{mean} =\frac{\beta(S/N)T_\mathrm{sys}}{G\sqrt{n_\mathrm{p}t_\mathrm{obs}\Delta{f}}}{\sqrt{\frac{\sigma}{1-\sigma}}}
	\label{eq:quadratic}
\end{equation}
where, $S_\mathrm{mean}$ is the mean pulse flux density, in units of mJy,
$\beta$ is a correction factor for FAST, in our case, correlation factor $\beta$ = 1,
S/N stands for the signal to noise ratio of pulsar, 
$T_\mathrm{sys}$ is the system noise temperature (K),
G is the gain of telescope (K Jy$^{-1}$). The system temperature is a function of the zenith angle $\theta_\mathrm{ZA}$, typically around 24 K, and is expressed as:
\begin{equation}
    T_\mathrm{sys} =\mathrm{P_0}\arctan(\sqrt{1+{\theta_\mathrm{ZA}}^\mathrm{n}}-\mathrm{P_1})+\mathrm{P_2}
	\label{eq:quadratic}
\end{equation}
where the range of $\theta_\mathrm{ZA}$ is 0$^\circ$ to 40$^\circ$. The values of $\mathrm{P_0}$, $\mathrm{P_1}$, $\mathrm{P_2}$, and n can be found in Table 4 of~\citet{2020RAA....20...64J}.
$n_\mathrm{p}$ is the polarization summed, 
observation duration is $t_\mathrm{obs}$, in units of seconds,
$\Delta{f}$ is observing bandwidth, in our case, the effective bandwidth is 400 MHz, 
$\sigma$ = ${n_\mathrm{on}}/{n_\mathrm{bin}}$ is the duty cycle.
We applied Eq. (\ref{eq:quadratic}) to estimate the flux density of pulsar at 1250 MHz.

\subsection{ Rotating vector model}
\label{sec:rvm} % used for referring to this section from elsewhere
Pulsar polarization observations could lead to a well-defined position angle (PA), and we used the \textbf{psrmodel} tool of the PSRCHIVE software package to fit the RVM \citep{1969ApL.....3..225R}. 
PA is described by the following formula:
\begin{equation}
    \mathrm{tan}(\psi_0-\psi) =\frac{\mathrm{sin}(\alpha)\mathrm{sin}(\phi_0-\phi)}{\mathrm{sin}(\zeta)\mathrm{cos}(\alpha)-\mathrm{cos}(\zeta)\mathrm{sin}(\alpha)\mathrm{cos}(\phi_{0}-\phi)}
	\label{eq:pa}
\end{equation}
where $\psi$ is the linear polarization position angle,
$\alpha$ is the magnetic inclination angle and $\phi$ is the pulsar rotational phase, 
$\zeta$ = $\alpha$ + $\beta$ (viewing angle),
$\psi_0$ and $\phi_0$ correspond to the position angle and rotational phase corresponding to the fiducial plane, respectively.

\section{RESULTS}
\label{sec:results} % used for referring to this section from elsewhere

\begin{figure*}
	\includegraphics[width=7in,height=6in]{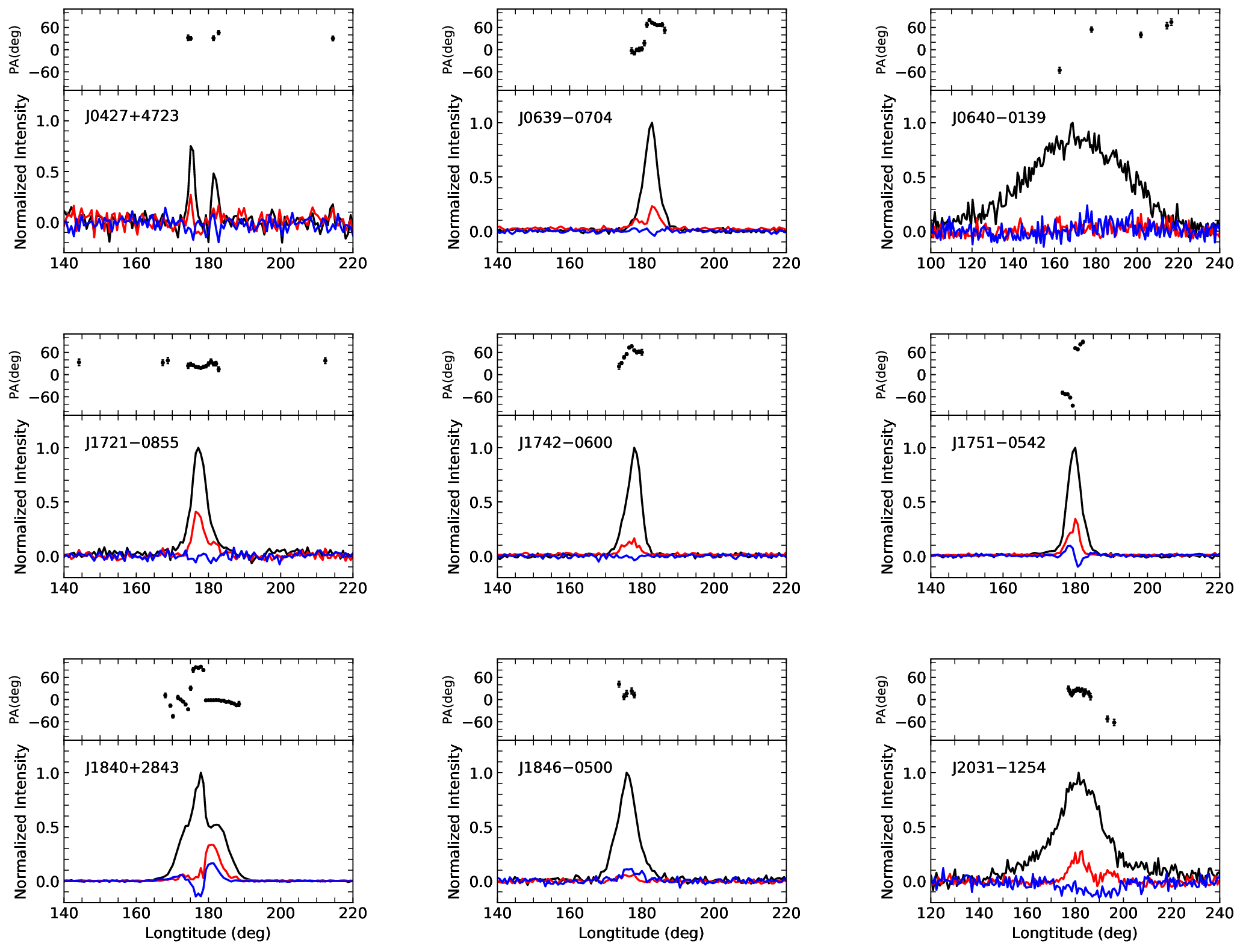}
    \caption{Integrated pulse profiles for nine pulsars, PSRs J0427+4723, J0639$-$0704, J0640$-$0139, J1721$-$0855, J1742$-$0600, J1751$-$0542, J1840+2843, J1846$-$0500, J2031$-$1254. The black line in the lower panel of each subplot represents the total intensity, and the red line and blue line represent the linear polarization intensity and circular polarization intensity respectively. From the plot, it can be seen that the peak intensity is normalized to 1.}
    \label{polzation}
\end{figure*}

In this work, we report the timing solutions for nine isolated normal pulsars, two isolated MSPs, and one MSP in a binary system. The sample we selected consists of pulsars discovered by CRAFTS that, at the time of writing this paper, have an observational span of over one year and have not yet any timing solutions published. Other CRAFTS pulsars were excluded due to either insufficient observational span or too few observations. The timing residuals are shown in  Figure~\ref{timing} with at least one year of follow-up timing observations.
It is worth noting that to ensure the TOA residuals $\chi^2$ are $\sim$ 1, we have scaled the TOA uncertainties by the so-called EFAC. These 12 pulsars timing solutions are reported in Tables~\ref{tab:t1} to \ref{tab:t5}, with JPL planetary ephemerides DE438\footnote{\url{ftp://ssd.jpl.nasa.gov/pub/eph/planets/bsp/de438.bsp}}, in TCB units.
We estimated the distance of the pulsar using YMW16 model \citep{2017ApJ...835...29Y}
and the NE2001 model \citep{2002astro.ph..7156C} 
for the Galactic distribution of free electrons.
We determine the spin-down luminosity ($\dot{E} = 4{\pi}^{2}I\dot{P}P^{-3}$),
the surface magnetic field ($B_\mathrm{surf} = 3.2\times10^{19}(P\dot{P})^{1/2}$) and the characteristic age ($\tau_\mathrm{c} = P/2\dot{P}$) from period and spin period dervative ($\dot{P}$). 

We selected the observation with the highest S/N and drew a polarization figure, as shown in Figure~\ref{polzation}.
The black solid lines represent the integrated pulse profiles with linearly polarized emission (in red) and circularly polarized emission (in blue),
respectively.
According to the RM measurement method described in Section~\ref{sec:obs}, we calibrated the pulse profiles and the results are
presented in Table~\ref{tab:rm}.
We measured the pulse width at 10$\%$ and 50$\%$ of the profile peak, denoted as $W_{10}$ and $W_{50}$ respectively.
Columns (5) to (7) in Table~\ref{tab:rm} give the fractional linear polarization L/S, the fractional net circular polarization V/S and the fractional absolute 
circular polarization $|\mathrm{V}|$/S.
According to the radiometer equation~\ref{eq:quadratic}, the flux density of the pulsar at 1.25 GHz is calculated and presented in the last column of 
Table~\ref{tab:rm}. The spin periods of the pulsars in this study range from 3.46 ms to 2.18 s, and their characteristic ages range from 1.43 Myr to 15.2 Gyr. We obtained the flux densities of these pulsars spanning from 16(6) $-$ 492(60) $\mu$Jy.

\subsection{ RVM fitting}

\begin{figure*}
\centering
	\includegraphics[width=5in,height=5in]{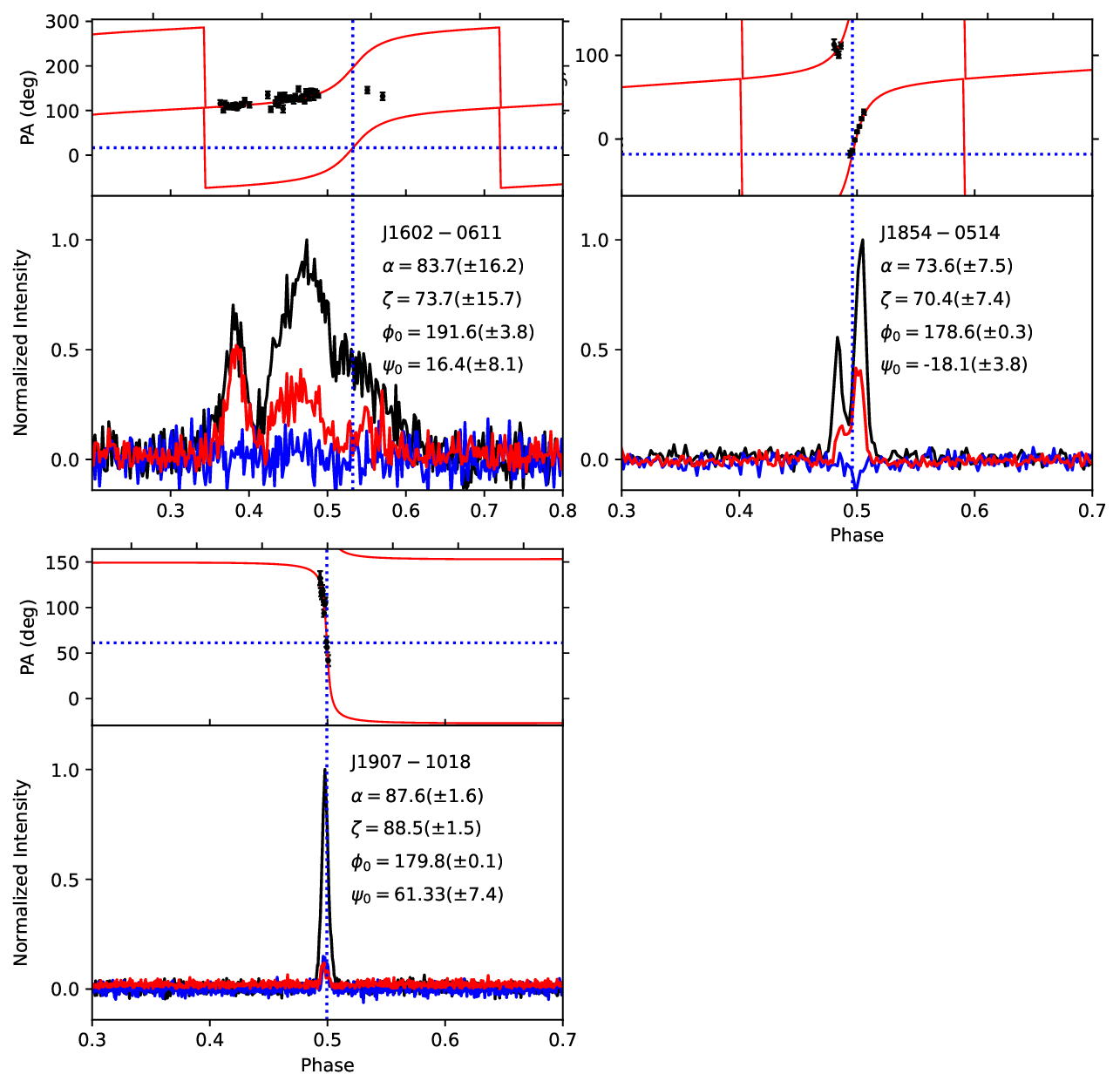}
    \caption{Integrated polarization profiles and view geometry of PSRs J1602$-$0611, J1854$-$0514 and J1907$-$1018. The points in the top panel of each subplot are the polarization position angles, and the errors are given. The red line on the upper panel represents the best fit result of the RVM model. Polarization position angle of four pulsars is $>$ 4 times the PA errors.
 }
    \label{rvm}
\end{figure*}
After calibration and RM fitting, only 3 in our 12 pulsars have sufficiently well-defined polarization angle (PA) curves.
Figure~\ref{rvm} shows the integrated polarization profile and viewing geometry obtained by fitting the RVM model.

PSR J1602$-$0611 is an MSP with a spin period of 3.95 ms, a DM of 27.626(1) cm$^{-3} $pc. From rmfit, we estimated the Faraday rotation measure of PSR J1602$-$0611 to be 5(3) rad m$^{-2}$, with the measured linear and circular polarizations being 38.2(2)$\%$ and 2(2)$\%$, respectively. Through the RVM fitting, we obtained the magnetic inclination angle, reference position angle, fiducial plane longitude and viewing angle with $\alpha$ = 84(16)$^\circ$, $\psi_0$ = 16(8)$^\circ$, $\phi_0$ = 192(4)$^\circ$ and $\zeta$ = 74(16)$^\circ$, respectively.
The estimated DM distance is 1.4 kpc and 1.9 kpc for NE2001 model and YMW16 model respectively.

%PSR J1751$-$0542 is a 2.00 s pulsar with the DM of 63.9(1) cm$^{-3} $pc. This pulsar has a Faraday rotation measure of $-$73(1) rad m$^{-2}$, the linear and circular polarization of this pulsar are 26.73$\%$  and  1.51$\%$, respectively. Through the RVM method, we obtained $\alpha$ = 145(567)$^\circ$, $\psi_0$ = $-$46(5)$^\circ$, $\phi_0$ = 179.1(2)$^\circ$ and $\zeta$ = 143(586)$^\circ$. This pulsar is beyond the traditional death line and will be discussed in detail in the Section~\ref{sec:deathline}.

PSR J1854$-$0514 is a pulsar with a spin period of 1.28 s and a DM of 281.1(2) cm$^{-3} $pc. From rmfit, it has a Faraday rotation measure of 178(1) rad m$^{-2}$. Through the RVM method, we measured $\alpha$ = 74(8)$^\circ$, $\psi_0$ = $-$18.1(4)$^\circ$, $\phi_0$ = 178.6(3)$^\circ$ and $\zeta$ = 70(7)$^\circ$.
The values obtained for the linear and circular polarization for this pulsar are {\bf 39(6)}$\%$ and {\bf $-$8(2)}$\%$, respectively. The inferred pulsar distance is 6.1 kpc and 8.7 kpc for the NE2001 and YMW16 model respectively. The spin-down luminosity and characteristic age of this pulsar are 3.5 $\times$ 10$^{31}$ erg s$^{-1}$ and 1.1 $\times$ 10$^7 $ yr respectively. PSR J1854$-$0514 exhibits a nulling phenomenon presented in Figure~\ref{single}.

PSR J1907$-$1018 is a 1.77 s pulsar with DM of 90.99(8) cm$^{-3} $pc.
 We obtained a Faraday rotation measure of 7(3) rad m$^{-2}$ though rmfit, this estimated RM is the lowest in our source list.
  Through the RVM method, we found
  $\alpha$ = 88(2)$^\circ$, $\psi_0$ = 61(7)$^\circ$, $\phi_0$ = 179.8(1)$^\circ$ and $\zeta$ = 89(2)$^\circ$. The estimated linear polarization and circular polarization of this pulsar are {\bf 12(2)}$\%$ and {\bf 11.20(3)}$\%$, respectively. The NE2001 model and YMW16 model predicted a DM distance of 2.7 kpc and 4.3 kpc, respectively.
 This pulsar belongs to relatively young pulsars with characteristic age of 3.79 $\times$ 10$^6$ yr and 
 has a spin-down luminosity of 5.3 $\times$ 10$^{31}$ erg s$^{-1}$. The single pulse sequence display pulse nulling phenomena for this pulsar (Figure~\ref{single}).

\begin{table*}
\begin{threeparttable}
\footnotesize
\caption{Best-fitting timing parameters for PSRs J0427$-$4723, J0639$-$0704, J0640$-$0139 and J1721$-$0855.}
\label{tab:t1}
\begin{tabular*}{\linewidth*}{lllll}
\hline
%\multicolumn{2}{c}{Fit and data-set} \\
Pulsar name &  PSR J0427+4723 & PSR J0639$-$0704 & PSR J0640$-$0139 & PSR J1721$-$0855\\
\hline\hline
$Measured$ $Parameters$ \\
Right ascension, $\alpha$ (hh:mm:ss)\dotfill &  04:27:51.465(7) & 06:39:58.644(10) &  06:40:45.280643(8) & 17:21:34.24(1)\\ 
Declination, $\delta$ (dd:mm:ss)\dotfill & +47:23:01.8(3)& $-$07:04:38.2(4) & $-$01:39:49.6639(3) & $-$08:55:08(1) \\ 
Pulse frequency, $\nu$ (s$^{-1}$)\dotfill & 0.463298405496(4) &  1.013699113957(3) & 288.892720892735(1) & 0.45843663480(2) \\ 
First derivative of pulse frequency, $\dot{\nu}$ (s$^{-2}$)\dotfill & $-$5.1263(3)$\times 10^{-15}$ & $-$1.67(3)$\times 10^{-16}$& $-$4.0879(5)$\times 10^{-15}$ & $-$6.59(1)$\times 10^{-16}$\\ 
Dispersion measure, DM(cm$^{-3}$ pc)\dotfill &54.0(2) & 105.06(8) & 149.983(2) & 46.9(5)\\

\hline
$Fitting$ $parameters$ \\
First TOA (MJD)\dotfill  &59192.7 & 59347.3 & 59486.0 & 59662.9\\
Last TOA (MJD)\dotfill &60127.0 & 60127.2 & 60186.1 & 60176.5 \\
Timing epoch (MJD)\dotfill & 59598.0 & 59737.0 & 59680.0 & 59668.0\\ 
Data span (yr)\dotfill & 2.56 & 2.14 & 1.92 & 1.41\\ 
Number of TOAs\dotfill & 23 & 26 & 38 & 19\\
Solar system ephemerides model\dotfill & DE438 & DE438 & DE438 & DE438\\
Rms timing residual ($\mu s$)\dotfill & 330.6 & 287.8 & 1.175 & 1359.6\\
TOA weighted factor, EFAC\dotfill &  1.1 &1.1 & 1.4 & 4.9\\ 

\hline
$Derived$ $parameters$ \\
Galactic longitude, $l$ ($^\circ$)\dotfill & 156.202 & 217.950 & 213.196 & 14.169\\
Galactic latitude, $b$ ($^\circ$)\dotfill & $-$1.017 & $-$5.758 & $-$3.135 & 15.385\\
DM distance, $d$ (kpc)\dotfill\\
\multicolumn{1}{r}{NE2001} &1.4 &  4.0 & 5.7 & 1.6\\
\multicolumn{1}{r}{YMW16} & 1.9 & 2.4 & 2.7  & 0.2\\
Characteristic age, $\tau_c$ (Myr) \dotfill & 1.43 & 96.34 & 1.12$\times10^{3}$ & 11.01\\
Surface magnetic field, $B_\mathrm{surf}$ ($10^{10}$G) \dotfill & 7.27$\times10^{2}$ & 40.50 & 4.17$\times10^{-2}$ & 2.65$\times10^{2}$\\
Spin-down luminosity, $\dot{E}$, ($10^{30}$erg s$^{-1}$) \dotfill & 9.4 & 6.7 & 4.7$\times10^{4}$ & 12\\
\hline

\end{tabular*}
\end{threeparttable}
\end{table*}

\begin{table*}
\begin{threeparttable}
\label{tab:t2}
\footnotesize
\caption{Best-fitting timing parameters for PSRs J1742$-$0559, J1751$-$0542, J1840+2843 and J1846$-$0500.}
\begin{tabular*}{\linewidth*}{lllll}
\hline
%\multicolumn{2}{c}{Fit and data-set} \\
Pulsar name  & PSR J1742$-$0559 & PSR J1751$-$0542 & PSR J1840+2843 & PSR J1846$-$0500\\
\hline\hline
$Measured$ $Parameters$ \\
Right ascension, $\alpha$ (hh:mm:ss)\dotfill  &  17:42:06.289(2)&17:51:32.606(7) &  18:40:29.589(10)&  18:46:26.917(8)\\ 
Declination, $\delta$ (dd:mm:ss)\dotfill & $-$05:59:57.79(11)  & $-$05:42:16.2(3)  & +28:43:13.7(3)& $-$05:00:19.5(9)\\ 
Pulse frequency, $\nu$ (s$^{-1}$)\dotfill  & 1.089112170924(3)& 0.499654292364(4) &1.457509613559(6)& 2.09803725694(3)\\ 
First derivative of pulse frequency, $\dot{\nu}$ (s$^{-2}$)\dotfill  & $-$7.408(2)$\times 10^{-16}$ & $-$5.75(5)$\times 10^{-17}$ & $-$2.2(5)$\times 10^{-17}$& $-$7.202(11)$\times 10^{-15}$\\
Dispersion measure, DM(cm$^{-3}$ pc)\dotfill  & 98.23(9) & 63.9(1) & 63.7(1)& 405.39(8)\\

\hline
$Fitting$ $parameters$ \\
First TOA (MJD)\dotfill  &59265.0 &59276.0 & 59246.0& 59423.6 \\
Last TOA (MJD)\dotfill  & 60130.6 & 60130.6 & 59971.1& 60069.9\\
Timing epoch (MJD)\dotfill  & 59618.0 &59631.0 &59608.0  & 59706.0\\ 
Data span (yr)\dotfill  &2.37 & 2.34 &1.99&1.77\\ 
Number of TOAs\dotfill  & 25 & 21 & 21&13 \\
Solar system ephemerides model\dotfill & DE438 & DE438 & DE438 & DE438\\
Rms timing residual ($\mu s$)\dotfill  &138.5 & 362.6 & 169.6&98.5\\
TOA weighted factor, EFAC\dotfill & 1.0 & 2.5 & 1.3& 1.0\\ 

\hline
$Derived$ $parameters$ \\
Galactic longitude, $l$ ($^\circ$)\dotfill  & 19.420 & 20.854 & 57.938& 27.909\\
Galactic latitude, $b$ ($^\circ$)\dotfill & 12.477 &10.573 & 14.855& $-$1.174\\
DM distance, $d$ (kpc)\dotfill\\
\multicolumn{1}{r}{NE2001}  & 3.1 &2.0 & 3.5& 6.8\\
\multicolumn{1}{r}{YMW16}  & 1.0 & 0.6 & 6.2& 6.5\\
Characteristic age, $\tau_c$ (Myr) \dotfill  & 23.30 & 1.38$\times10^{2}$ & 1.05$\times10^{3}$ & 4.62\\
Surface magnetic field, $B_\mathrm{surf}$ ($10^{10}$G) \dotfill  & 76.64 & 68.7 & 8.52 & 89.4\\
Spin-down luminosity, $\dot{E}$, ($10^{30}$erg s$^{-1}$) \dotfill  & 32 & 1.1 & 1.3 & 6.0$\times10^{2}$ \\
\hline

\end{tabular*}
\end{threeparttable}
\end{table*}

\begin{table*}
\begin{threeparttable}
\footnotesize
\caption{Best-fitting timing parameters for PSRs J1854$-$0514, J1907$-$1018 and J2031$-$1254.}
\label{tab:t3}
\begin{tabular*}{\linewidth*}{llll}
\hline
%\multicolumn{2}{c}{Fit and data-set} \\
Pulsar name & PSR J1854$-$0514 & PSR J1907$-$1018  & PSR J2031$-$1254\\
\hline\hline
$Measured$ $Parameters$ \\
Right ascension, $\alpha$ (hh:mm:ss)\dotfill  &  18:54:18.408(6) &  19:07:42.088(6) &   20:31:30.35440(4)\\ 
Declination, $\delta$ (dd:mm:ss)\dotfill & $-$05:14:26.1(5) & $-$10:18:42.5(4) & $-$12:54:28.826(3)\\ 
Pulse frequency, $\nu$ (s$^{-1}$)\dotfill  & 0.781281237119(6) &  0.566323304509(2) & 172.760569821455(4)\\ 
First derivative of pulse frequency, $\dot{\nu}$ (s$^{-2}$)\dotfill  & $-$1.1253(6)$\times 10^{-15}$ & $-$2.3686(5)$\times 10^{-15}$ & $-$1.796(6)$\times 10^{-16}$\\
Dispersion measure, DM(cm$^{-3}$ pc)\dotfill  & 281.1(2) & 90.99(8) & 22.938(1)\\

\hline
$Fitting$ $parameters$ \\
First TOA (MJD)\dotfill  & 59314.0 & 59423.6 & 58971.9\\
Last TOA (MJD)\dotfill  & 60044.9 & 60221.5 & 60129.8\\
Timing epoch (MJD)\dotfill   & 59640.0 & 59793.0 &59478.0\\ 
Data span (yr)\dotfill & 2.00 & 2.18 & 3.17\\ 
Number of TOAs\dotfill  & 22 & 15 & 17\\
Solar system ephemerides model\dotfill & DE438 & DE438  & DE438\\
Rms timing residual ($\mu s$)\dotfill   & 435.9 & 251.7 &1.394\\
TOA weighted factor, EFAC\dotfill  & 3.0 & 4.8 &1.4\\ 

\hline
$Derived$ $parameters$ \\
Galactic longitude, $l$ ($^\circ$)\dotfill  &28.590 & 25.529  & 32.228\\
Galactic latitude, $b$ ($^\circ$)\dotfill  & $-$3.023 & $-$8.263 & $-$27.951\\
DM distance, $d$ (kpc)\dotfill\\
\multicolumn{1}{r}{NE2001}  & 6.1 & 2.7 &1.1\\
\multicolumn{1}{r}{YMW16}  & 8.7 & 4.3 & 1.4\\
Characteristic age, $\tau_c$ (Myr) \dotfill   & 11.0 & 3.79 & 1.52$\times10^{4}$\\
Surface magnetic field, $B_\mathrm{surf}$ ($10^{10}$G) \dotfill  & 1.55$\times10^{2}$ & 3.65$\times10^{2}$ & 1.89$\times 10^{-2}$\\
Spin-down luminosity, $\dot{E}$, ($10^{30}$erg s$^{-1}$) \dotfill  & 35 & 53 & 1.2$\times10^{3}$\\
\hline

\end{tabular*}
\end{threeparttable}
\end{table*}

\begin{table}
\begin{threeparttable}
\footnotesize
\caption{Best-fitting timing parameters for PSR J1602$-$0611.}
\label{tab:t5}
\begin{tabular*}{\linewidth}{ll}
\hline
%\multicolumn{2}{c}{Fit and data-set} \\
Pulsar name &  PSR J1602$-$0611\\
\hline\hline
$Measured$ $Parameters$ \\
Right ascension, $\alpha$ (hh:mm:ss)\dotfill &  16:02:17.24172(5)\\ 
Declination, $\delta$ (dd:mm:ss)\dotfill & $-$06:11:43.475(2)\\ 
Pulse frequency, $\nu$ (s$^{-1}$)\dotfill &  253.14146512594(1)\\ 
First derivative of pulse frequency, $\dot{\nu}$ (s$^{-2}$)\dotfill &  $-$3.03(1)$\times 10^{-16}$\\
Dispersion measure, DM(cm$^{-3}$ pc)\dotfill  & 27.626(1)\\
Orbital period, $P_b$ (d)\dotfill & 18.23999206(2) \\ 
Projected semi-major axis of orbit, $x$ (lt-s)\dotfill & 12.179114(1) \\ 
TASC (MJD)\dotfill & 59805.5271999(3) \\ 
EPS1\dotfill & $-$1.94(2)$\times 10^{-5}$ \\ 
EPS2\dotfill & 7.6(3)$\times 10^{-6}$ \\
Binary model\dotfill & ELL1 \\

\hline
$Fitting$ $parameters$ \\
First TOA (MJD)\dotfill & 59357.7\\
Last TOA (MJD)\dotfill & 60187.4\\
Timing epoch (MJD)\dotfill  &59805.5\\ 
Data span (yr)\dotfill & 2.27\\ 
Number of TOAs\dotfill & 55\\
Solar system ephemerides model\dotfill & DE438 \\
Rms timing residual ($\mu s$)\dotfill  &5.3\\
TOA weighted factor, EFAC\dotfill & 3.7\\ 

\hline
$Derived$ $parameters$ \\
Galactic longitude, $l$ ($^\circ$)\dotfill  & 4.327\\
Galactic latitude, $b$ ($^\circ$)\dotfill & 33.102\\
DM distance, $d$ (kpc)\dotfill\\
\multicolumn{1}{r}{NE2001}  &1.4\\
\multicolumn{1}{r}{YMW16} & 1.9\\
Characteristic age, $\tau_c$ (Myr) \dotfill  & 1.32$\times 10^{4}$\\
Surface magnetic field, $B_\mathrm{surf}$ ($10^{10}$G) \dotfill & 1.38$\times 10^{-2}$\\
Spin-down luminosity, $\dot{E}$, ($10^{30}$erg s$^{-1}$) \dotfill  & 3.0$\times 10^{3}$\\
Inferred eccentricity, $e$ (10$^{-5}$)\dotfill & 2.08(2)\\
Mass function, $f$ (10$^{-3}$)$M_{\odot}$\dotfill &5.8(3)\\
Minimum companion mass, $m_{c,min}$($M_{\odot}$)\dotfill &0.2458\\
Median companion mass, $m_{c,med}$($M_{\odot}$)\dotfill & 0.2889\\
\hline

\end{tabular*}
\end{threeparttable}
\end{table}

\begin{table*}
	\centering
	\caption{Pulse width at 50 $\%$ and 10$\%$ of the profile peak, rotation measure (RM), polarization parameters (L/S, V/S and $|\mathrm{V}|$/S respectively) and flux density.}
	\label{tab:rm}
	\begin{tabular}{lcccccccr} % four columns, alignment for each
		\hline
		PSR & P &$W_{10}$ & $W_{50}$ & RM & L/S & V/S & $|\mathrm{V}|$/S  & $S_\mathrm{1.25GHz}$\\
              & (s)  & (ms) & (ms) & (rad $m^{-2}$) & (\%) & (\%) & (\%) & ($\mu$Jy) \\
		\hline
		J0427+4723 & 2.16 &26.4(3) & 22.0(3) & 431(21) & 12.5(2) & $-$25(8) & 37(8) & 16(6)\\
		J0639$-$0704 & 0.99 &28.5(3) & 12.7(3) & 10(2) & 18(1) & 1.1(2) & 4.6(2) & 100(4)\\
		J0640$-$0139 & 3.46$\times10^{-3}$ &3.31(4) & 1.60(4) & 168(16) & 3(1) & 2.2(3) & 12.5(3) & 189(14)\\
%        J1115$-$0958 & 26.78 & 5.22 & 45(1) & 42.80 & 0.17 & 7.29 & 283(36)\\
		J1602$-$0611 & 3.95$\times10^{-3}$ &2.68(5) & 1.48(5) & 5(3) & 38.2(2) & 2(2) & 8(2) & 97(3)\\
		J1721$-$0855 & 2.18 &28.5(8) & 12.7(8) & 10(3) & 34(2) & $-$2(1) & 6(1) & 216(19)\\
  		J1742$-$0559 & 0.92 &25.1(1) & 11.6(1) & 75(1) & 16(4) & $-$1.5(9) & 2.6(9) & 222(19)\\
		J1751$-$0542 & 2.00 &24.6(4) & 11.9(4) & 73(1) & 27(3) & 1.5(5) & 9.1(5) & 258(31)\\
		J1840+2843 & 0.69 &52.8(1) & 28.5(1) &75(1) & 22(1) & 5.3(5) & 13.8(5) & 492(60)\\
        J1846$-$0500 & 0.48 & 33.4(1) & 15.1(1) & 381(18) & 4.4(6) & 10.1(2) & 10.3(3) & 99(7)\\
        J1854$-$0514 & 1.28 & 33.5(5) & 10.5(5) & 178(1) & 39(6) & $-$8(2) & 10(2) & 90(10)\\
        J1907$-$1018 & 1.77 &14.9(1) & 6.9(1) & 7(3) & 12(2) & 11.20(3) & 11.20(3) & 95(19)\\
%		J2006$-$1313 & 17.28 & 8.23 & 26(1) & 23.73 & 21.67 & 33.58 & 94(11)\\
		J2031$-$1254 & 5.79$\times10^{-3}$ & 2.10(2) & 0.56(2) & $-$7(4) & 13.1(8) & $-$12.8(2) & 12.8(2) & 132(12)\\
		\hline
	\end{tabular}
\end{table*}

\subsection{Nulling}
\label{nulling pulsars}

\begin{figure}
	\includegraphics[width=\columnwidth]{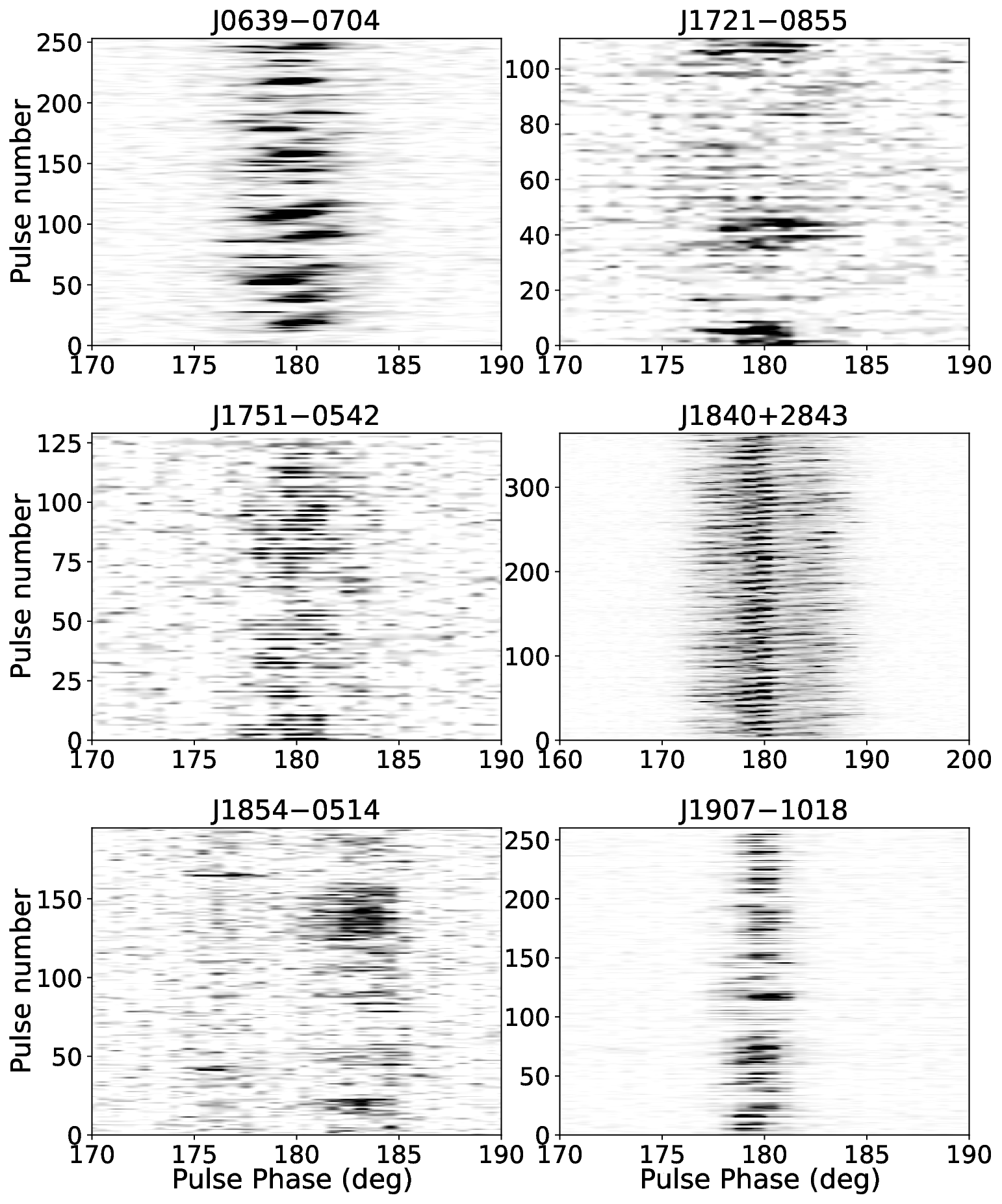}
    \caption{Single pulse sequences display pulse nulling phenomena for five new discovered PSRs J0639$-$0704, J1721$-$0855, J1751$-$0542, J1854$-$0514 and J1907$-$1018. PSR J1840+2843 shows the subpulse drift phenomenon.}
    \label{single}
\end{figure}

\begin{figure}
	\includegraphics[width=\columnwidth]{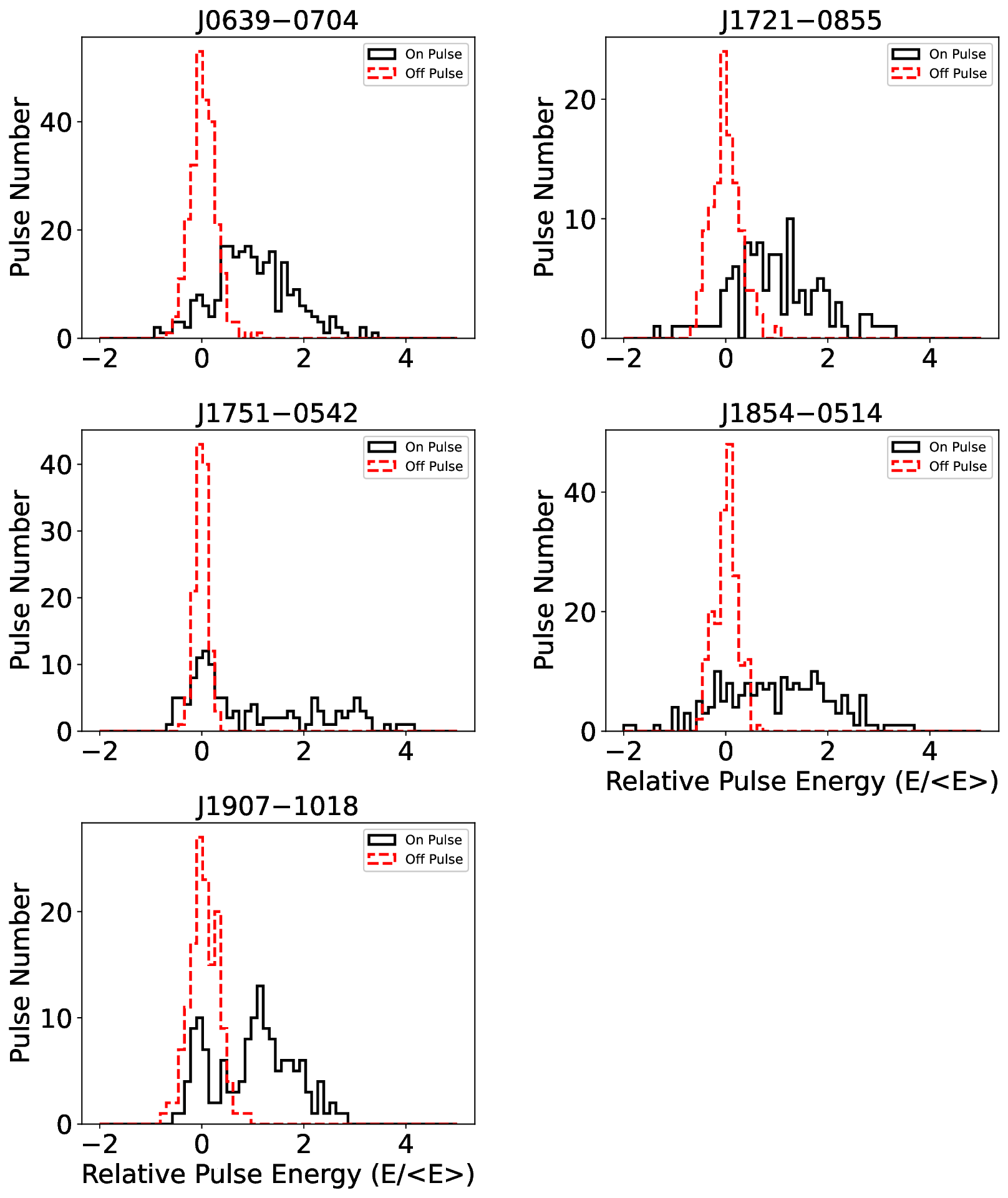}
    \caption{Pulse energy distributions for the on-pulse (black solid histogram) and off-pulse (red dashed histogram) regions for five pulsars. The energies are normalised by the mean on-pulse energy.
    }
    \label{energy}
\end{figure}

In our work, five pulsars display the phenomenon of pulse nulling, they are all normal pulsars, and their single pulse sequences are shown in Figure~\ref{single}. The pulse energy distributions for the on-pulse and off-pulse windows are shown in Figure~\ref{energy}. The on-pulse window was defined as the total longitude range over which significant pulsed emission is seen.
The off-pulse energy was measured in an equal length of on-pulse bins. 
As can be seen from Figure~\ref{energy}, the on-pulse energy of each pulsars has a peak adjacent to zero.
Following studies by \citet{1976MNRAS.176..249R}, 
we calculate the nulling fraction (NF) by subtracting a scaled version of the off-pulse histogram from the
on-pulse histogram until the sum of the difference counts in bins with E $<$ 0 was zero.
The uncertainty of NF is given by $\sqrt{n_\mathrm{p}}/N$, where $n_\mathrm{p}$ is the number of null pulses and N is the total number of pulses \citep{2007MNRAS.377.1383W}.
We choose observations with high signal-to-noise ratio to obtain NF and then calculate the average value.
The average NF values (expressed as NF$_\mathrm{R}$) for the five pulsars are listed in the fifth column of Table~\ref{NF}.

Recent studies have found that \citet{1976MNRAS.176..249R} tends to overestimate the NF in its algorithm, especially for pulsars with weaker emission, leading to significant biases~\citep{2018ApJ...855...14K,2023ApJ...948...32A}. Here, we employ a more robust mixture model method described by \citet{2023ApJ...948...32A} to recalculate the NF (denoted as NF$_\mathrm{A}$) for these five nulling pulsars, as shown in the last column of Table~\ref{NF}. Our calculations use the same data as previously used for calculating the NF. The results indicate that \citet{1976MNRAS.176..249R} may indeed overestimate the NF for some pulsars. However, it's worth noting that our effective exposure time per observation is only 4 minutes, which might introduce biases. This warrants further verification of their NF through longer observations in the future.

\begin{table}
	\centering
	\caption{The average nulling fractions for five pulsars.  $T_\mathrm{obs}$, $N_\mathrm{tot}$ and $N_\mathrm{null}$ represent the duration, the total number of pulses and number of null pulses of observation, respectively. The NF$_\mathrm{R}$ and NF$_\mathrm{A}$ represent the NF calculated using the methods of \citet{1976MNRAS.176..249R} and \citet{2023ApJ...948...32A}, respectively.}
	\label{NF}
	\begin{tabular}{ccccc|c} % four columns, alignment for each
		\hline
		PSR & $T_\mathrm{obs}$ & $N_\mathrm{tot}$ & $N_\mathrm{null}$ & NF$_\mathrm{R}$ & NF$_\mathrm{A}$\\
                 & (min) & &  &\% &\%\\
		\hline
		J0639$-$0704 & 52.8 & 3220 & 1196 & 37(4) & 8(1)\\
		J1721$-$0855 & 36.6 & 1012 & 335 & 31(5) & 14(4)\\
		J1751$-$0542 & 37.2 & 1123 & 444 &  45(6) & 24(3)\\
        J1854$-$0514 & 36.0 & 1692 & 546 & 33(4) &4(2)\\
        J1907$-$1018 & 44.6 & 1524 & 489 & 30(5) &16(2)\\
		\hline
	\end{tabular}
\end{table}

PSR J0639$-$0704 is a pulsar with a spin period of 0.98 s and a DM of 105.06 cm$^{-3} $pc. 
Through rmfit, we find a RM of 10(2) rad m$^{-2}$ for this pulsar. The measured values of linear and circular polarization are 18(1)$\%$  and  1.1(2)$\%$, respectively.
With a DM of 105.06(8) cm$^{-3} $pc, the NE2001 model and YWM16 model predict DM distance of 4.0 kpc and 2.4 kpc respectively. It has a low-$\dot{E}$ of 6.7 $\times$ 10$^{30}$ erg s$^{-1}$ with a characteristic age of 9.63 $\times$ 10$^{7}$ yr.
The pulse sequence of PSR 0639$-$0704 shows a nulling phenomenon (Figure~\ref{single}). The single pulse sequence of this pulsar indicates that it may exhibit subpulse drifting phenomenon. Currently, due to the short duration of each observation, it is challenging to effectively analyze whether there is subpulse drifting behavior.

PSR J1721$-$0855 is a 2.18 s pulsar.
The measured linear and circular polarizations of this pulsar are 34(2)$\%$ and $-$2(1)$\%$, respectively 
based on Faraday rotation measure of 10(3) rad m$^{-2}$.
The distance based on the pulsar' DM of 46.9(5) cm$^{-3} $pc is 1.6 kpc and 0.2 kpc for the NE2001 and YMW16 model respectively.
The estimated DM distance is the nearest for this pulsar in our source list.
With characteristic age of 1.1 $\times$ 10$^7 $ yr,
PSR J1721$-$0855 is an old pulsar having $\dot{E}$ of  1.2 $\times$ 10$^{31}$ erg s$^{-1}$ and this pulsar shows a nulling phenomenon presented in Figure~\ref{single}.

\begin{figure}
	\includegraphics[width=\columnwidth]{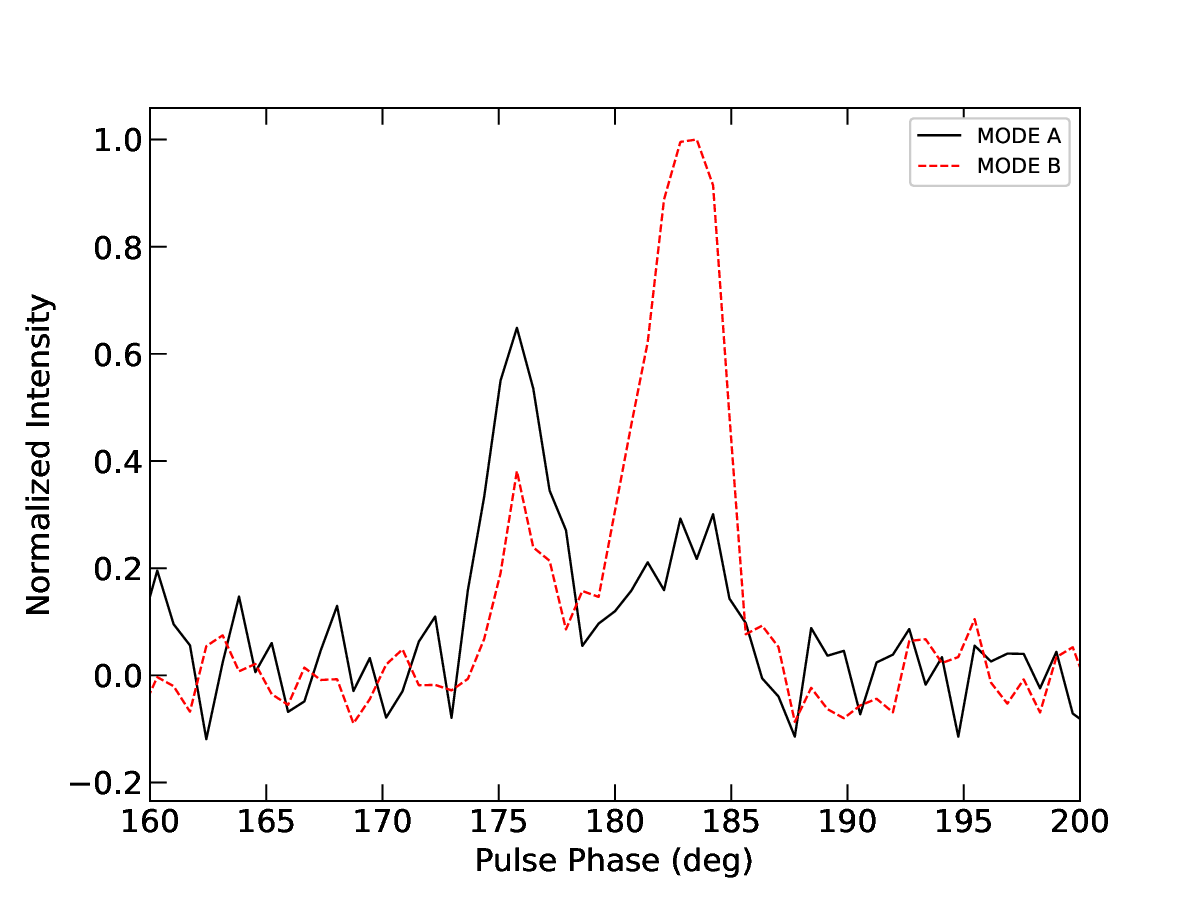}
    \caption{The mean pulse profiles of the two distinct modes of PSR J1854$-$0514.
    }
    \label{modeab}
\end{figure}

From Figure~\ref{single}, we observe that PSR J1854$-$0514 not only exhibits pulsar nulling phenomena but also demonstrates mode-changing characteristics. This is a double-peaked profile pulsar. To differentiate between the two distinct modes of this pulsar, we define the minimum intensity between two adjacent components as the boundary separating the double-peaked profile. When the peak pulse intensity on the left exceeds that on the right, we define it as Mode A. Conversely, it is defined as Mode B. We have stacked a total of 36 minutes of observations of this pulsar and present the mean pulse profiles of its two distinct modes in Figure~\ref{modeab}. Mode A and Mode B account for approximately 25$\%$ and 75$\%$ of the total data, respectively.

\subsection{Subpulse drifting}
\label{subpulse drifting pulsar}

\begin{figure}
	\includegraphics[width=3in,angle=270]{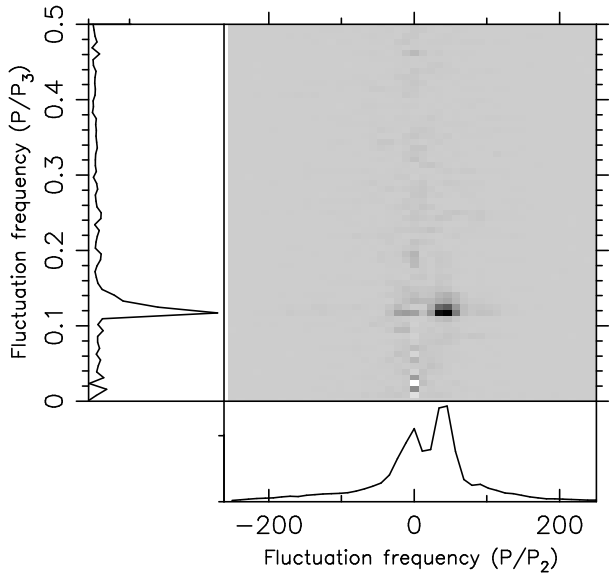}
    \caption{The 2DFS for the PSR J1840+2843. The left panel of the figure shows significant peaks between 0.1246 and 0.1254 cycles per period, indicating that the pulsar exhibits strong periodic intensity variations. The lower panel of the figure is asymmetric at the 0 axis, suggesting the presence of drifting phenomena in this pulsar.
    }
    \label{2DFS}
\end{figure}

PSR J1840+2843 shows an apparently periodic subpulse drifting phenomenon. We plotted the two-dimensional fluctuation spectrum (2DFS) \citep{2002A&A...393..733E} in Figure~\ref{2DFS}.
The frequency axis of 2DFS corresponds to $P$/$P_3$, where $P$ is the pulsar rotation period and $P_3$ represents the periodicity of subpulse drifting.
The horizontal frequency axis of 2DFS corresponds to $P$/$P_2$, where $P_2$ represents the characteristic horizontal time separation between drifting bands.

In Figure~\ref{2DFS}, the side panel of the 2DFS shows a clear spectral peak occurring between 0.1246 and 0.1254 cycles per period (cpp).
This spectral feature corresponds to the characteristic $P_3 = 7.99 \pm 0.02P$.
There are two peaks in the lower panel of 2DFS, with the peak on the right corresponding to the horizontal drifting period of drift bands.
This spectral feature corresponds to the characteristic $P_2 = 9.25^{\circ} \pm 0.04^{\circ}$.

 \begin{figure}
	\includegraphics[width=\columnwidth]{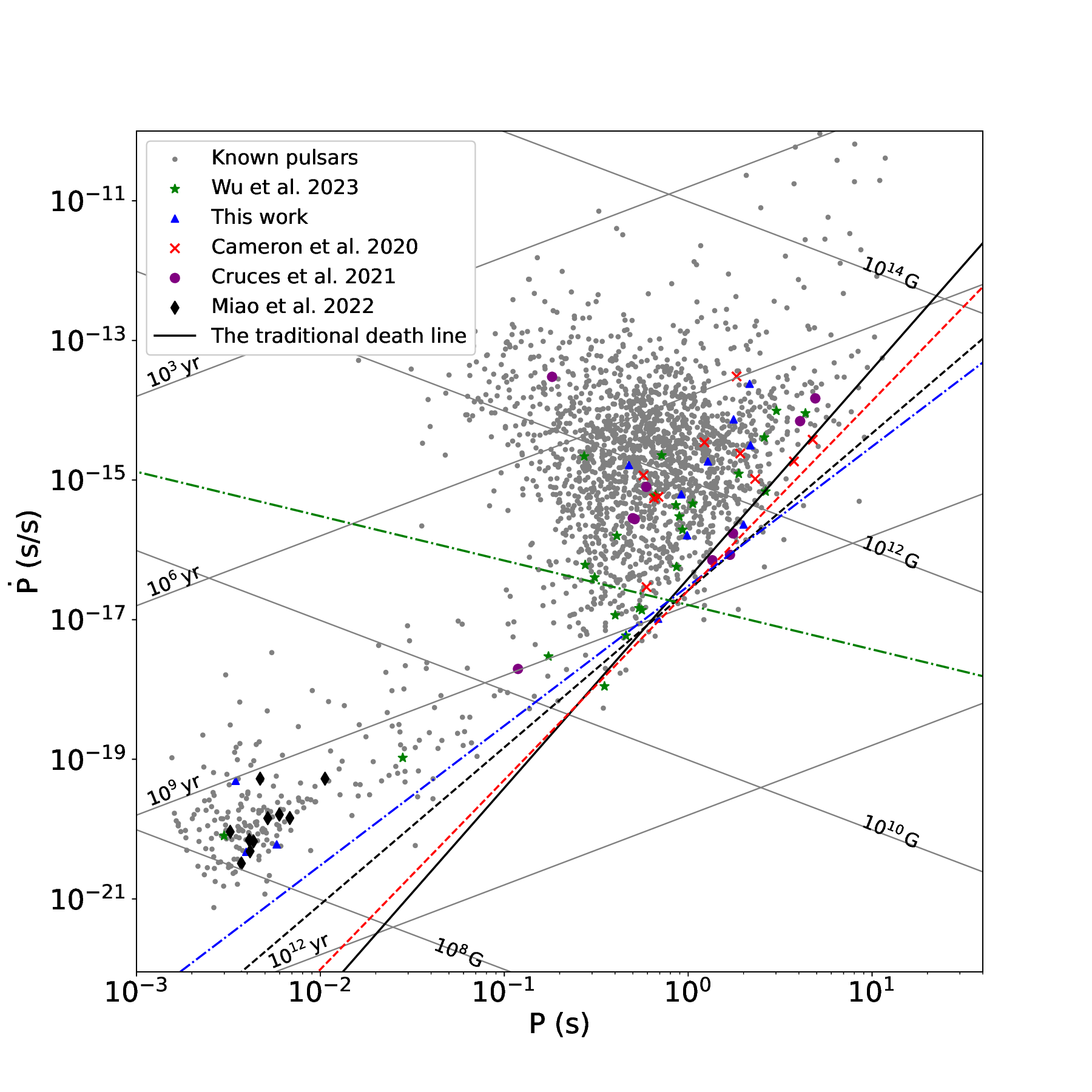}
    \caption{A $P - \dot{P}$ diagram of known pulsars and new pulsars presented in this work. The grey dots are known pulsars.
    The red stars represent pulsars reported in \citet{2023MNRAS.522.5152W}. The red ``x" represents pulsars reported in \citet{2020MNRAS.495.3515C}.
    The red circle represents pulsars reported in \citet{2021MNRAS.508..300C}. The red rhombus represents pulsars reported in \citet{2023MNRAS.518.1672M}.
    The blue triangle represents pulsars reported in this work.
    We used several colored lines to plot the death line under different death line models,
    pulsars below the death lines are not expected to emit radio signals.
    The traditional death line \citep{1992A&A...254..198B} is the black line.
    Formula (9) in  \citet{1993ApJ...402..264C} corresponds to a red dashed death line.
    The black dashed line and blue dashed line are based on curvature radiation from the vacuum gap model and the space-charged-limited flow (SCLF) model 
    as proposed by \citet{2000ApJ...531L.135Z}.
    The green dashed line is based on SCLF models for inverse Compton scattering. }

    \label{ppdot}
\end{figure}

\subsection{The traditional death line}
\label{sec:deathline}
Figure~\ref{ppdot} displays  69 newly discovered pulsars from CRAFTS on a $P$-$\dot{P}$ diagram, compared with the known pulsars listed in the PSRCAT (version 1.70).
These 69  new pulsars consist of 11 pulsars reported in \citet{2020MNRAS.495.3515C}, 
10 pulsars reported in \citet{2021MNRAS.508..300C},
12 MSPs reported in \citet{2023MNRAS.518.1672M},
24 pulsars reported in \citet{2023MNRAS.522.5152W}
and 12 pulsars reported in this work.

According to \citet{1992A&A...254..198B}, the expression for the traditional death line :
\begin{equation}
    (B_\mathrm{s}/10^{12})/{P^2} = 0.17
	\label{eq:death line}
\end{equation}
which corresponds to a spin-down luminosity ($\dot{E}$) of 1.53 $\times$$10^{30}$ erg s$^{-1}$.
When the spin-down luminosity is lower than 1.53 $\times$$10^{30}$ erg s$^{-1}$, 
the pulsar exceeds the traditional death line in the $P - \dot{P}$ diagram.
In this work, two pulsars (PSRs J1751$-$0542 and J1840+2843) are located below the traditional death line in the $P - \dot{P}$ diagram.

%PSR J1115$-$0956 was first discovered by the Parks Radio Telescope and its timing solution was released \citep{2020MNRAS.496.4836S}. We used FAST observations to discover that PSR J1115$-$0956 is a nulling pulsar, and updated the ephemerides and provided more detailed discussions. PSR J1115$-$0956 has period of 1.31s and DM of 16.1(1) cm$^{-3} $pc. With a Faraday rotation measure of 45(1) rad m$^{-2}$, we measured 42.80 $\%$  and  0.17 $\%$ of the linear and circular polarization respectively. The NE2001 model predicted a DM distance of 0.68 kpc and YMW16 model predicted a DM distance of 1.05 kpc. The characteristic age of this pulsar is 6.33 $\times$ 10$^8 $ yr, which belongs to relatively old pulsar. It has a low-$\dot{E}$ 5.7 $\times$ 10 $^{29}$ erg/s, which is beyond the traditional death line. Only 20 pulsars have $\dot{E}$ small than this pulsar's spin-down luminosity in the ATNF pulsar catalogue (version:1.70).  It is very interesting that the single pulse sequence of this pulsar shows the phenomenon of nulling (Fig~\ref{single}).

PSR J1751$-$0542 is a 2.00 s pulsar with a DM of 63.9(1) cm$^{-3} $pc. This pulsar has a Faraday rotation measure of 73(1) rad m$^{-2}$, the linear and circular polarizations of this pulsar are 27(3)$\%$ and 1.5(5)$\%$, respectively. PSR J1751$-$0542 has a spin-down luminosity of 1.1 $\times$ 10$^{30}$ erg s$^{-1}$ and characteristic age of 1.38 $\times$ 10$^8 $ yr.
Only 40 pulsars have $\dot{E}$ smaller than this pulsar. 
%Compared with J1115$-$0956, this pulsar is relatively younger.
%Similar to the PSR J1115$-$0956, 
Single pulse sequence of this pulsar shows nulling phenomenon (Figure~\ref{single}).
The NE2001 model predicted a DM distance of 2.0 kpc while YMW16 model predicted a  DM distance of 0.6 kpc.

PSR J1840+2843 has a period of 0.69 s and a characteristic age of 1.05 $\times$ 10$^9 $ yr and a spin-down luminosity of 1.3 $\times$ 10$^{30}$ erg s$^{-1}$, below the traditional death line.
Only 46 pulsars have $\dot{E}$ smaller than this pulsar. This pulsar shows subpulse drifting phenomenon as shown in Figure~\ref{single}. It has a DM of 63.7(1) cm$^{-3} $pc and a Faraday rotation measure of 75(1) rad m$^{-2}$. 
The linear and circular polarizations are 22(1)$\%$ and 5.3(5)$\%$, respectively.
The estimated pulsar distance is 3.5 kpc and 6.2 kpc for the NE2001 and YMW16 model respectively.

%One pulsar is adjacent to traditional death line in our sample. PSR J2006$-$1313 has a spin period of 0.37 s and a low DM of 24.60(1) cm$^{-3} $pc. The spin-down luminosity of this pulsar is 2.4 $\times$ 10$^{30}$ erg s$^{-1}$. It is an old pulsar with a characteristic age of 1.9 $\times$ 10$^9 $ yr. From rmfit, the estimated Faraday rotation measure of 26(1) rad m$^{-2}$ for PSR J2006$-$1313. With a RM, the estimated linear and circular polarization of this pulsar are 23.73$\%$  and  21.67$\%$, respectively. The NE2001 model and YMW16 model predicted a DM distance of 1.10 kpc and 1.24 kpc, respectively.

\subsection{Millisecond pulsars}
\label{sec:msp}

\begin{figure}
	\includegraphics[width=\columnwidth]{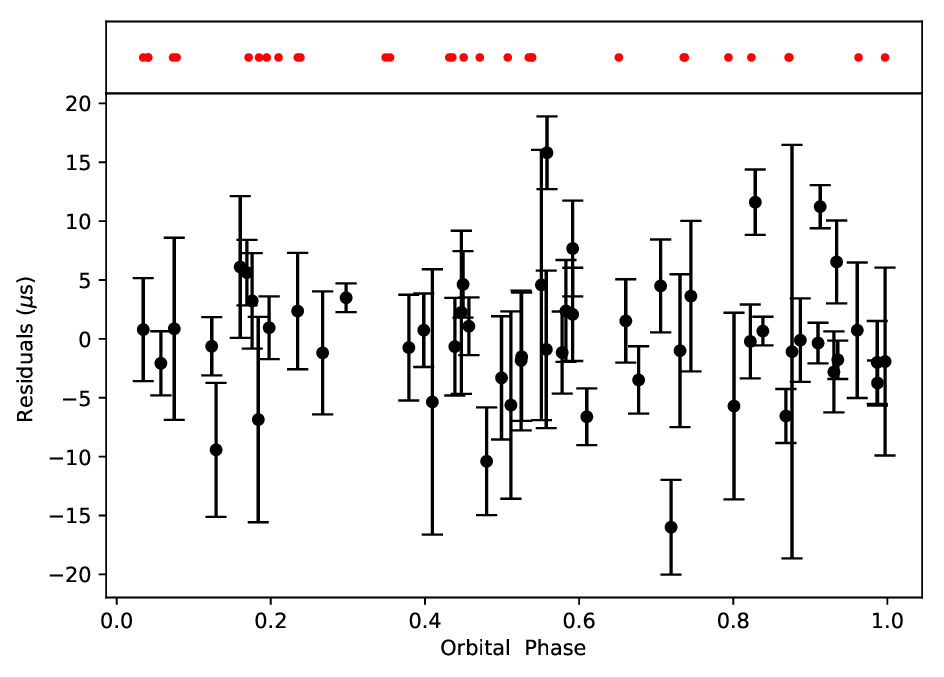}
    \caption{Timing residuals with binary orbit phase for PSR J1602$-$0611, based on the timing solution presented in Table~\ref{tab:t5}. The red dots represent observations without pulse signals. In order to see them more clearly, we have placed them in the upper subplot.
    }
    \label{binphase}
\end{figure}

\begin{figure}
	\includegraphics[width=\columnwidth]{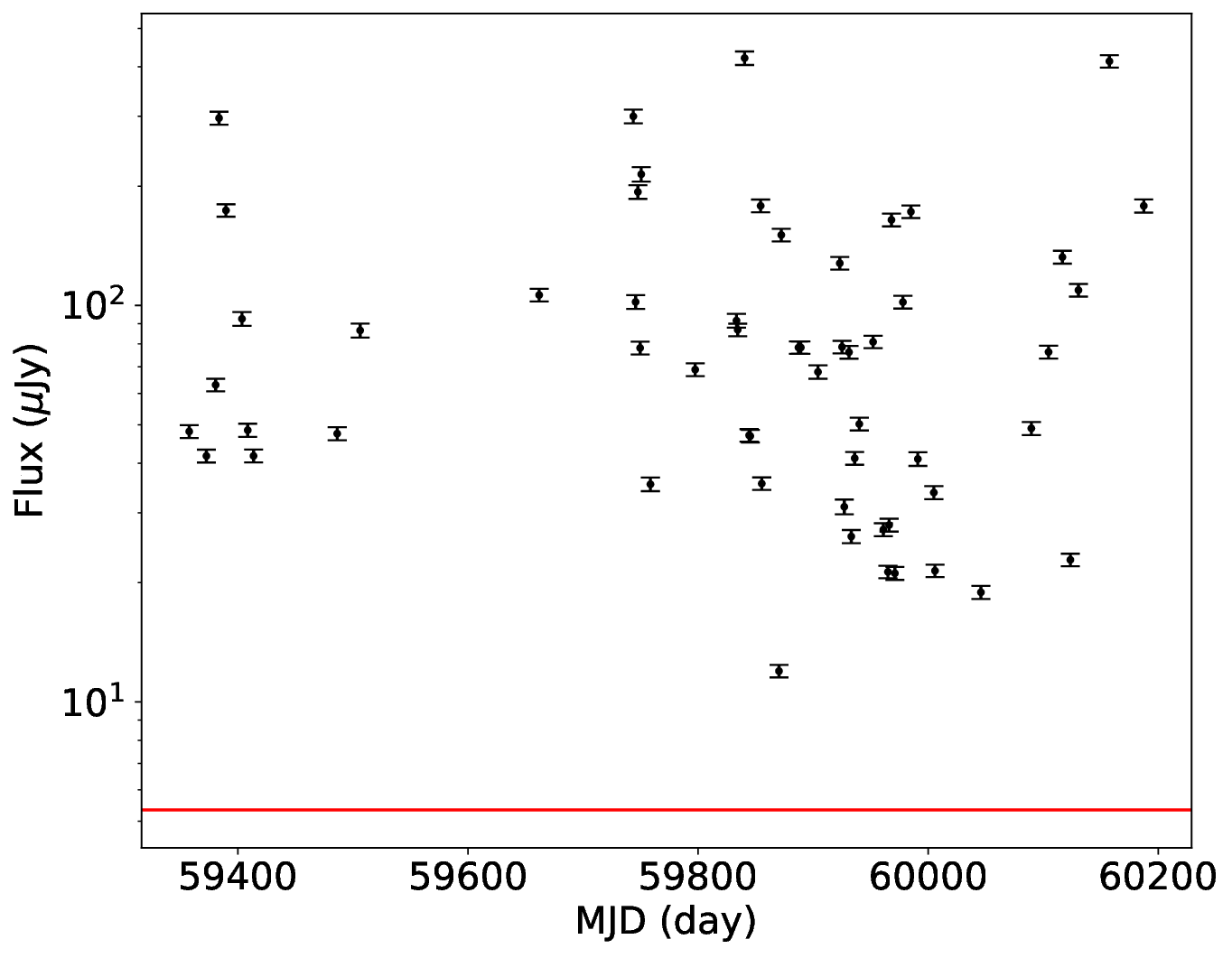}
    \caption{Flux density of PSR J1602$-$0611. The red solid line represents the flux density limit for FAST observation for 4 minutes.
    }
    \label{flux}
\end{figure}
In this work, a total of three MSPs are reported, two of which are isolated MSPs (PSRs J0640$-$0139 and J2031$-$1254) and one is in a binary system (PSR J1602$-$0611).

The rotation period of PSR J0640$-$0139 is 3.46 ms, with a stable spin-down rate of 4.8981(6) $\times$10$^{-20}$\,s\,s$^{-1}$.
From \textbf{rmfit}, it has a Faraday rotation measure of 168(16) rad m$^{-2}$.
The linear and circular polarizations for PSR J0640$-$0139 were measured to be low with 3(1)$\%$  and 2.2(3)$\%$, respectively.
With a DM of 149.983(2) cm$^{-3} $pc, 
the NE2001 model and YWM16 model predicts a distance of 5.7 kpc and 2.7 kpc respectively.
%It is an old pulsar with a characteristic age of 1.12 $\times$ 10$^{9}$ yr.
It is an energetic pulsar which has spin-down luminosity of 4.7 $\times$10$^{34}$ erg s$^{-1}$, 
exceeds the energy loss rate threshold of  3 $\times 10^{34}$ erg s$^{-1}$ for potential gamma ray pulsar \citep{1999ApJ...516..297T}.
We checked the observation source\footnote{\url{https://heasarc.gsfc.nasa.gov/cgi-bin/W3Browse/}} of Fermi Gamma-ray Space Telescope, but no gamma ray counterpart is found for this pulsar. 

PSR J1602$-$0611 is an MSP with a spin period of 3.95 ms, a DM of 27.626(1) cm$^{-3} $pc.
It is in an 18.24 day binary orbit with an eccentricity of 2.03 $\times$ 10$^{-5}$.
This pulsar has not detected signals in many observations (the red dots in Figure~\ref{binphase}). We used Eq.~(\ref{eq:quadratic}) to calculate the flux density of all visible signals of PSR J1602$-$0611 (Figure~\ref{flux}), and it can be seen that the flux density of PSR J1602$-$0611 varies immensely, with the ratio of the highest value to the lowest on being 35.2. 
Among them, the one with the lowest flux density is 11.96 $\mu$Jy. It is close to the flux density limit of FAST observation for 4 minutes ($\sim$5.34 $\mu$Jy) (The red solid line in Figure~\ref{flux}). 
Therefore, we suspect that the non-detection may be attributed to interstellar scintillation or variations in the pulsar's flux density due to changes in the particle flow from the pulsar's magnetosphere. We used the corresponding Python script in package dynsp$\textunderscore$acf~\footnote{\url{https://github.com/larskuenkel/dynsp_acf}} to estimate the scintillation timescale of the source, which is 2.35(2) minutes. As the exposure time is short with four minute each epoch, the causes of non-detection and flux variations need further investigations. To investigate whether there is an eclipse in PSR J1602$-$0611, we show in Figure~\ref{binphase} the timing residuals are plotted against the orbital phase.
No noticeable trend is seen.
This pulsar has a wide pulse profile. The pulse widths $W_{10}$ and $W_{50}$ are 2.6841 ms and 1.4834 ms, respectively.
Assuming that the mass of the pulsar is 1.4 $M_{\odot}$, the mass function of the system is 0.0058(3) $M_{\odot}$ and 
the minimum mass of the companion star is 0.2458 $M_{\odot}$. 
%The $\tau_c$ of this pulsar is 1.32 $\times$ 10$^{10} $ yr.
According to the orbital parameters of this pulsar and the minimum mass of companion star, the companion is most likely a Helium white dwarf (He - WD).

%It seems that these observations are not concentrated on one of the orbital phase, but rather evenly distributed, implying that the interstellar scintillation timescale is greater than the observation length. The dominant cause of these variations can be assessed using the methods proposed by~\citep{2023MNRAS.520.1311W}.

 PSR J2031$-$1254 is an isolated millisecond pulsar with a spin period of 5.79 ms and a DM of 22.938(1) cm$^{-3} $pc.
 This pulsar has a stable spin-down rate of 6.02(2) $\times$10$^{-21}$\,s\,s$^{-1}$ and Faraday rotation measure of $-$7(4) rad m$^{-2}$. The estimated linear and circular polarization of this pulsar are 13.1(8)$\%$ and $-$12.8(2)$\%$, respectively. The distance inferred based on the pulsar' DM of 22.938(1) cm$^{-3} $pc is 1.1 kpc and 1.4 kpc for the NE2001 and YMW16 model respectively.
This pulsar has spin-down luminosity of 1.2 $\times$ 10$^{33}$ erg s$^{-1}$.

\subsection{The remaining pulsars}
\label{remain}
PSR J0427+4723 is a sharp double-peaked pulsar with a spin period of 2.15 s. 
The left component of the pulse profile is dominant.
This pulsar has a high Faraday rotation measure of 431(21) rad m$^{-2}$, and the linear and circular polarizations for PSR J0427+4723 were measured to be 12.5(2)$\%$ and $-$25(8)$\%$, respectively.
With a DM of 54.0(2) cm$^{-3} $pc , the NE2001 model and YWM16 model predicts DM distance of 1.4 kpc and 1.9 kpc respectively.
The characteristic age of this pulsar is  1.43 $\times 10^6 $ yr, indicating that it is a relatively young pulsar.

PSR J1742$-$0559 has a period of 0.92 s. 
We find a RM of 75(1) rad m$^{-2}$ though rmfit for this pulsar,
the measured values of the linear and circular polarization are 16(4)$\%$ and $-$1.5(9)$\%$, respectively.
With a DM of 98.23(9) cm$^{-3} $pc ,
the obtained  pulsar' distance is 3.1 kpc and 1.0 kpc for the NE2001 and YMW16 model respectively.
The spin-down luminosity and characteristic age of this pulsar are 3.2 $\times$ 10$^{31}$ erg s$^{-1}$ and 2.33 $\times$ 10$^7 $ yr respectively.

PSR J1846$-$0500 is a pulsar with a period of 0.48 s.
This pulsar has higher Faraday rotation measure of 381(18) rad m$^{-2}$.
The obtained linear and circular polarizations of this pulsar are 4.4(6)$\%$ and 10.1(2)$\%$, respectively.
This pulsar has the highest DM of 405.39(8) cm$^{-3} $pc in our sample. The NE2001 model and YMW16 model predicted a DM distance of 6.8 kpc and 6.5 kpc, respectively.
This pulsar has a spin-down luminosity of 6.0 $\times$ 10$^{32}$ erg s$^{-1}$ and 
 belongs to relatively young pulsars with characteristic age of 4.62 $\times$ 10$^6$ yr.

\section{DISCUSSION}
\label{sec:conclusion}
We reported the timing results of twelve new pulsars from the CRAFTS survey. These pulsars have spin periods ranging from 3.46 ms to 2.18 s and their characteristic ages range from 1.43Myr to 15.2Gyr. We obtained the flux densities of these pulsars spanning from 16(6) $-$ 492(60) $\mu$Jy. All the twelve pulsars are not associated with Galactic Supernova Remnants\footnote{\url{http://www.mrao.cam.ac.uk/surveys/snrs/}} \citep{2019JApA...40...36G}. We reported new pulsars' the phase-connected timing ephemerides, polarization profiles, and Faraday rotation measurements. Nine out of twelve pulsars are normal pulsars, three out of twelve are millisecond pulsars. Among these pulsars, five ones shows the phenomenon of pulse nulling, and one pulsar shows subpulse drift phenomena (Figure~\ref{single}). In our work, PSR J0640$-$0139 and PSR J2031$-$1254 are isolated MSPs. %Neutron stars in the low mass X-ray binaries were spun up to millisecond pulsars by accreting mass and angular momentum from the companion star \citep{1991PhR...203....1B}. Excluding those MSPs in Globular clusters, which may be formed through dynamic processes, all MSPs should be in binary systems. However, approximately 20$\%$ of MSPs are isolated in the galactic field. \citet{1988Natur.334..227V} and \cite{1988Natur.334..225K} proposed that the donor stars may have been ablated by the $\gamma$-ray and energetic particles emitted by the MSPs. \citet{2013ApJ...775...27C} study suggests that the evaporation timescale may be too long unless a very high evaporation efficiency ($\sim$0.1). \citet{2020A&A...633A..45J}  proposed a new process to form isolated Millisecond pulsar. When the central density of NS rises above the quark deconstrained density, a phase transition (PT) from NSs to strange stars (SS) may occur, and a suitable kick could disrupt the binary. At present, the predicted NS-SS PT event has not been confirmed through observation.

\begin{table}
	\centering
	\caption{The average values of $P$ and $\dot{P}$ for pulsars below the traditional death line (A) and normal pulsars (B). The definition of normal pulsars is pulsars with a spin period greater than 30 ms.}
	\label{paverage}
	\begin{tabular}{ccc} % four columns, alignment for each
		\hline
	   	& $P$ (s)& $\dot{P}$ (\,s\,s$^{-1}$) \\
		\hline
		A & 3.125 & $2.630\times10^{-15} $\\
        B &0.887 & $4.823\times10^{-13}$ \\
		\hline
	\end{tabular}
\end{table}

\subsection{Testing radiation theory}
%From expression~\ref{eq:death line}, when the spin-down luminosity of pulsar is lower than 1.53 $\times$$10^{30}$ erg s$^{-1}$, the pulsars exceed the traditional death line.
%In our sample, two pulsars, PSRs J1751$-$0542 and J1840+2843, are located below this line.
Out of the 3400 pulsars in the pulsar population, there are 62 ones below the death line, which corresponds to $\sim$1.9$\%$ of the entire population.
The periods of these 62 ones span from 0.3456 to 75.89 second, with period derivatives between 0.00054 $\times 10^{-18}$ -- $255 \times 10^{-15}$\,s\,s$^{-1}$, 
ages of between 3.24  $\times 10^{6}$ -- 10.1 $\times 10^{9}$ yr.

A series of Parkes surveys using multibeam receivers have discovered $\sim$1261 pulsars so far, of which only 27 pulsars are below the death line (data is from ATNF version 1.70), indicating a 2.2 percent occurrence. 
At present, as for the 72 CRAFTS pulsars with spin-down rate reported, 10 (including two from this work) of which are located below the traditional death line, implying a ratio of 14$\%$. 
These comparisons suggest that FAST can found more pulsars below the traditional death line, 
which means that FAST has higher sensitivity.

We calculated the average values of $P$ and $\dot{P}$ for pulsars below the traditional death line and normal pulsars (Table~\ref{paverage}). It can be seen that the average $\dot{P}$ of pulsars below the traditional death line is two orders of magnitude smaller than the average $\dot{P}$ of normal pulsars. Additionally, their periods are relatively large, indicating that pulsars below the traditional death line have older ages.

PSR J1751$-$0542 exhibits pulse nulling phenomenon, and PSR J1840+2843 shows subpulse drift phenomenon. Furthermore, it should be noted that not all pulsars below the death line are nulling pulsars.

Figure~\ref{ppdot} shows death lines for various models. Each death line corresponds to a point where the radio emission from a pulsar is expected to be turned off due to insufficient electron pairs. The black solid line represents the traditional dead line model, where inadequate potential in the polar cap leads to radio-quiet \citep{1975ApJ...196...51R,1992A&A...254..198B}. The red dashed line, referred to as the \cite{1993ApJ...402..264C} model, considers complex magnetic structures with highly curved field lines. The black dashed line and blue dashed line correspond to the vacuum gap model and the space-charge-limited flow model, respectively~\citep{2000ApJ...531L.135Z}. These models were obtained by considering curvature radiation and inverse Compton scattering under the general relativistic frame-dragging effect~\citep{1979ApJ...231..854A}.
%Radio radiation disappears when gaps cannot form above the polar cap area. By incorporating general relativistic frame dragging \citep{1979ApJ...231..854A} and considering curvature radiation and inverse Compton scattering, the \citet{2000ApJ...531L.135Z} proposed the vacuum gap model (black dashed line) and space-charge-limited-flow model (blue dashed line) in Figure~\ref{ppdot}. 

PSR J0901$-$4046 has a spin period of 75.88 s, making it the longest known pulsar in terms of spin period \citep{2022NatAs...6..828C}. It is located above the death line of the space-charge-limited flow radio-emission model. This indicates that in the presence of a multipolar magnetic field configuration, non-relativistic charges can flow from the polar cap. PSR J1751$-$0542 can also be explained using this model, but PSR J1840+2843 is located below this model. It is hoped that in the future, more low-$\dot{E}$ pulsars can be found to test radiation theory.

%It is worth noting that PSR J2006$-$1313 is a pulsar adjacent the traditional death line, but located below death line of space-charge-limited flow radio-emission model.

%\cite{2014ApJ...784...59S} gives another explanation for the process behind the death line of pulsar emission.
%They believed that radio emission is considered to have the highest possible efficiency $\zeta \equiv L/\dot{E}$ $\approx$ 0.01.
%The radio efficiency of PSR J1115$-$0956 is $\zeta \approx$ 0.00314.
%We suggested that PSR J1115$-$0956 can be explained using the observation limit line proposed by \citet{2020RAA....20..188W}.
%This observation limit line corresponds to the maximum possible minimum-$\dot{E}$ of $\approx$ $10^{28}$ erg/s
%under the conditions of 1kpc and 10$\mu$Jy.

%According to the observation limit line theory, the more sensitive the telescope is, the more it can measure low-$\dot{E}$ pulsars. It is hoped that in the future, thanks to the high sensitivity of FAST, more low-$\dot{E}$ pulsars can be found to test radiation theory.

\begin{figure}
	\includegraphics[width=\columnwidth]{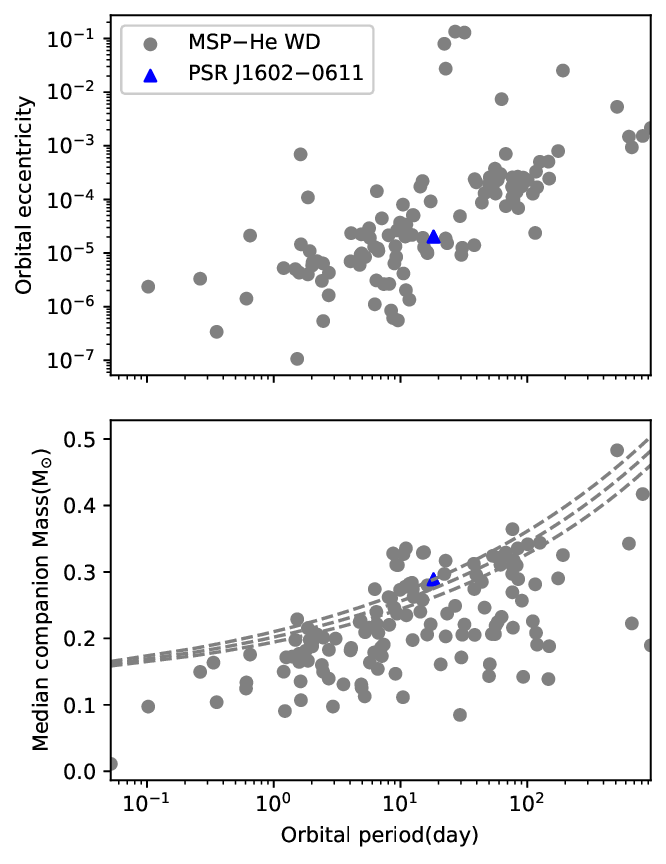}
    \caption{The MSPs with with light white dwarf (WD) companions from PSRCAT (version 1.70). MSP - He WD systems are shown with grey dots.
    Top panel: The correlation between orbital period and eccentricity of MSP and light WD companion systems.
    Bottom panel:  The correlation between orbital period and companion mass of MSP and light WD companion systems.
    The dashed lines are the numerical calculations of the MSP - He WD system by \citet{1999A&A...350..928T}. The median companion mass estimation is given based on their mass functions assuming an orbital inclination of $i=60^\circ$ and $M_\mathrm{psr}$ = 1.4 $M_\odot$.}.
    \label{bp}
\end{figure}

\subsection{The possible MSP - He WD binary system}
\label{binary}
MSPs is spin-up through accreting matter and angular momentum from companion stars~\citep{1991PhR...203....1B}.
During the transition to MSP, the magnetic field strength will decrease (usually from $10^{12}$ G to $10^8$ G) due to aging  \citep{1992ApJ...395..250G} or accretion \citep{1974SvA....18..217B}. The surface magnetic field strength of PSR J1602$-$0611 is 1.38 $\times 10^8$ G.
Assuming that the mass of the pulsar is 1.4 $M_\odot$ and the orbital inclination $i$ is $60^{\circ}$, according to the mass function,
the median companion mass of PSR J1602$-$0611 is 0.2889 $M_\odot$.
\citet{1999A&A...350..928T} proposed a companion mass prediction for Millisecond pulsar and Helium white dwarf systems (Figure~\ref{bp}, bottom panel). 
\citet{1999A&A...350..928T} predicted the correlation between orbital and companion mass for MSP - He WD system, as depicted in the bottom panel of Figure~\ref{bp} with black dashed lines.
We found that the orbital period and the companion mass distribution of  PSR J1602$-$0611 are consistent with MSP - He WD systems.
The distribution of orbital period and eccentricity of PSR J1602$-$0611 also conforms to MSP - He WD systems (Figure~\ref{bp}, top panel).
Therefore, we believe that this pulsar belongs to the MSP - He WD systems.
MSP - He WD systems in globular cluster are not considered in our sample, because the complex situation in globular cluster may make the eccentricity the other stars in this system inaccurate.

\begin{figure}
	\includegraphics[width=\columnwidth]{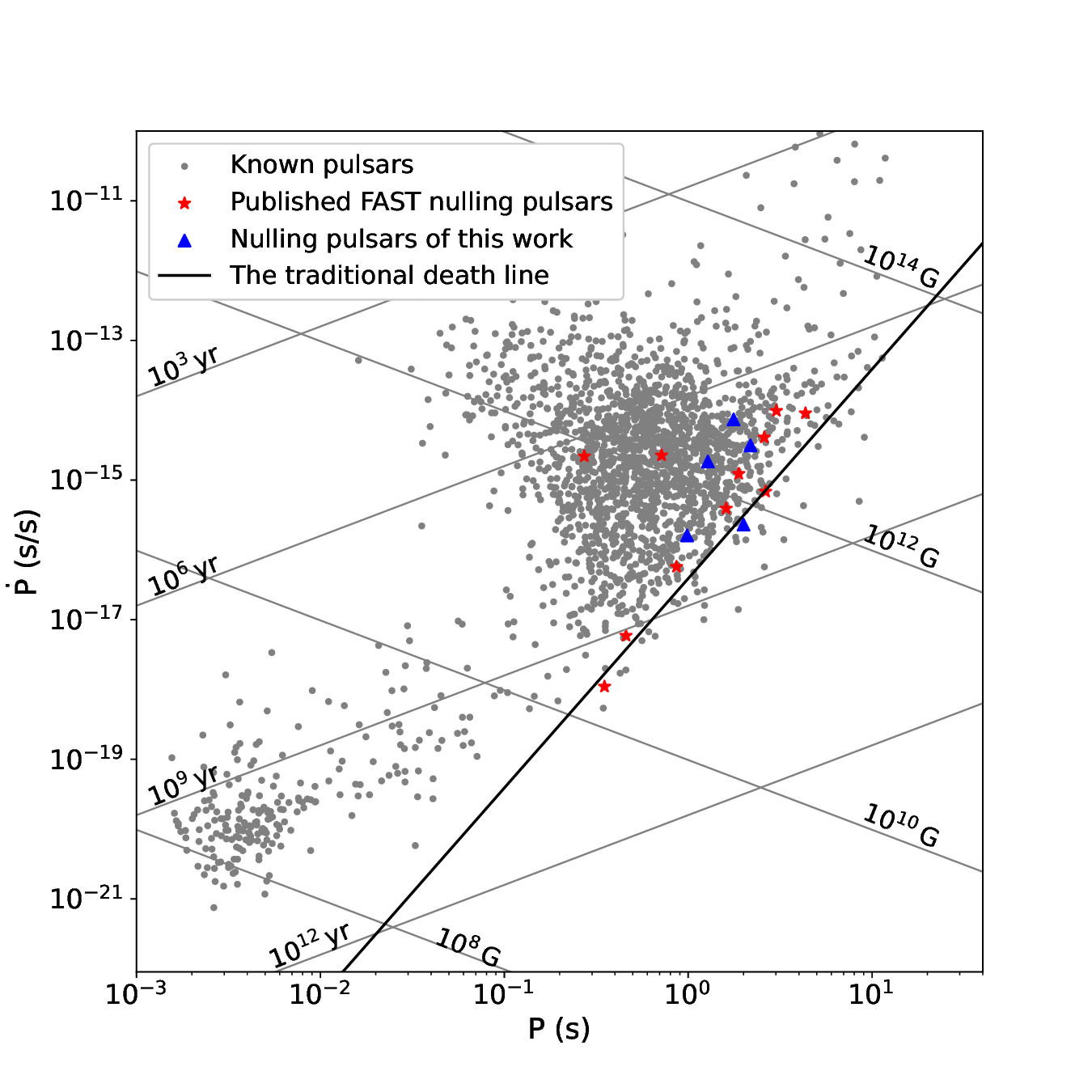}
    \caption{A $P - \dot{P}$ diagram of known pulsars and FAST's newly discovered nulling pulsars. The grey dots are known pulsars. The red star represents the published nulling pulsars discovered by FAST. The nulling pulsars in our study are represented by the blue triangles.}.
    \label{ppnulling}
\end{figure}

\begin{figure}
	\includegraphics[width=\columnwidth]{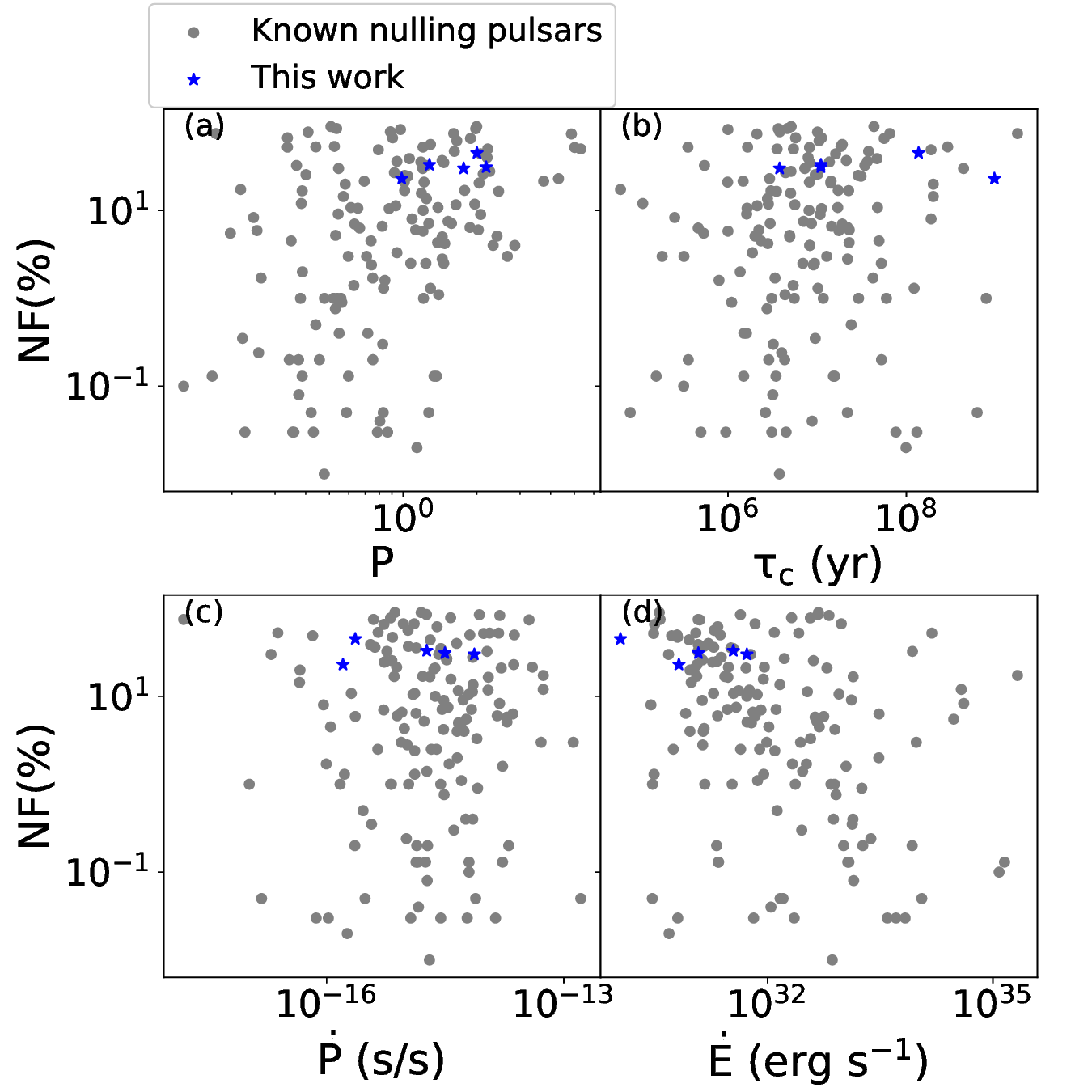}
    \caption{The correlations between NF and pulsar period, $\tau_c $, $\dot{P}$ and $\dot{E}$. The gray dots in the figure come from \citet{2020A&A...644A..73W}.}.
    \label{nf}
\end{figure}

\subsection{Nulling phenomenon}
\label{null}
Although recent results suggest that the method proposed by \citet{1976MNRAS.176..249R} may overestimate the NF values, this needs to be validated with larger samples in the future. Since most prior research has used~\citet{1976MNRAS.176..249R} method to estimate NF, for consistency, we still use the NF obtained from~\citet{1976MNRAS.176..249R} method in this study to compare with the NF of known nulling pulsars and provide a brief discussion.

From the normalized single-pulse energy distribution, we calculated the NF for these pulsars, with the largest one being around 45$\%$.
In Figure~\ref{ppnulling}, we plot the $P - \dot{P}$ diagram of the newly discovered nulling pulsars by FAST.
It seems that the new nulling pulsars discovered by FAST are often long period or old pulsars, which are near or even exceed the traditional death line. This means they all have relatively lower $\dot{E}$.
From Figure~\ref{ppdot} and Figure~\ref{ppnulling}, we can see that not all of the old or long-period pulsars found by FAST have pulse nulling phenomenon. This suggests that not all pulsars near even beyond the traditional death line will exhibit the nulling behavior, that FAST's high sensitivity enables the detection of more low $\dot{E}$ pulsars and facilitates testing of the radiation theory.

In order to compare the distribution of nulling pulsars in our work with known nulling pulsars, we plotted the relationship between nulling fraction and period, $\tau_c $, $\dot{P}$ and $\dot{E}$  of nulling pulsars in Figure~\ref{nf}. It can be seen that the nulling pulsars in our work has a larger period and larger NF as shown in panel (a). And panel (b) shows that the nulling pulsars in our work are older in age. The distribution of $\dot{P}$ is relatively uniform (panel (c)). Moreover, it can be seen that the nulling pulsars we found have lower $\dot{E}$, and one nulling pulsar in our work has the lowest $\dot{E}$ among these known nulling pulsars. Often, pulsars with older age have larger NFs, while pulsars with larger $\dot{E}$ tend to have smaller NFs. The relationship between long periods and large NF is the most significant \citep{1992ApJ...394..574B,2020A&A...644A..73W}, and the nulling pulsars in our study also confirm these relationships.

\subsection{ Flux density and luminosity}
It is difficult to determine the luminosity of a pulsar, but we can calculate the monochromatic luminosity \citep{2012hpa..book.....L,2014ApJ...784...59S} based on the formula:
\begin{equation}
   L_v \equiv S_{v}d^2
	\label{eq:luminosity}
\end{equation}
Where $S_v$ is the mean flux density measured at different frequencies.

\begin{figure*}
	\includegraphics[width=\textwidth]{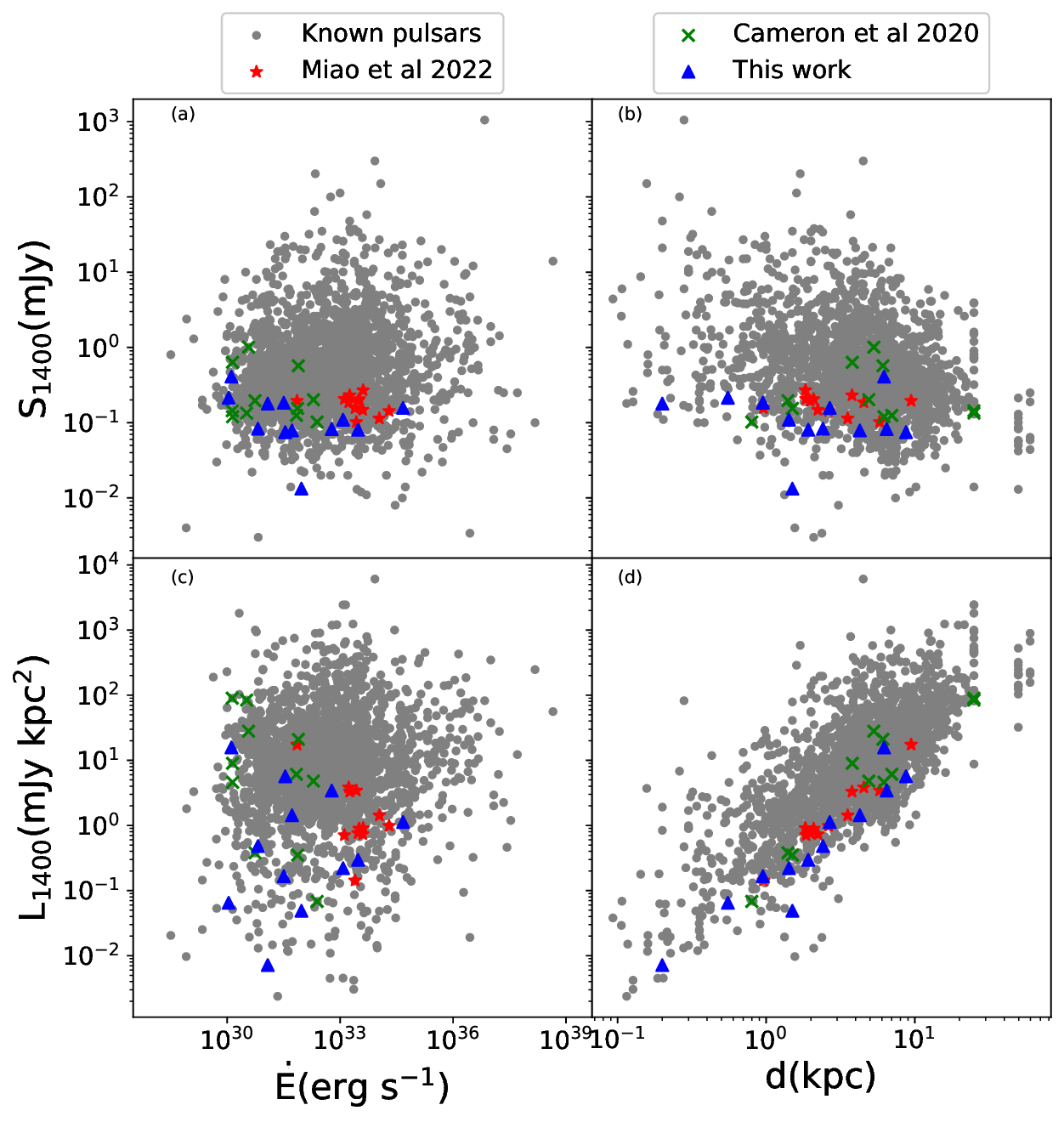}
    \caption{Panel (a) and (c) show the distributions of the mean flux density and the radio luminosity ($L_{1400}$) on spin-down luminosity ($\dot{E}$), respectively. The dependence of the mean flux density and the radio luminosity on distance from YMW16. The blue triangles represent pulsars reported in this work.  The green ``x" stand for CRAFTS pulsars in \citet{2020MNRAS.495.3515C}, and red stars are CRAFTS millisecond pulsars in \citet{2023MNRAS.518.1672M}. Other known pulsars are represented by gray dots, and the data is from ATNF (version 1.70).}
    \label{se}
\end{figure*}

In order to compare its average flux density with that of known pulsars, 
their flux densities at 1400 MHz are calculated using a pulsar spectral index. In this study, we utilize the average spectral index of $-$1.64 from \citet{2024arXiv240410542X} to calculate the flux density of pulsars at 1400 MHz.
%Since the average spectral index of pulsars in ATNF is $\sim$ $-$1.8225, we used a spectral index of $-$1.82 to calculate the flux density of pulsar at 1400 MHz in our work.

\begin{figure}
	\includegraphics[width=\columnwidth]{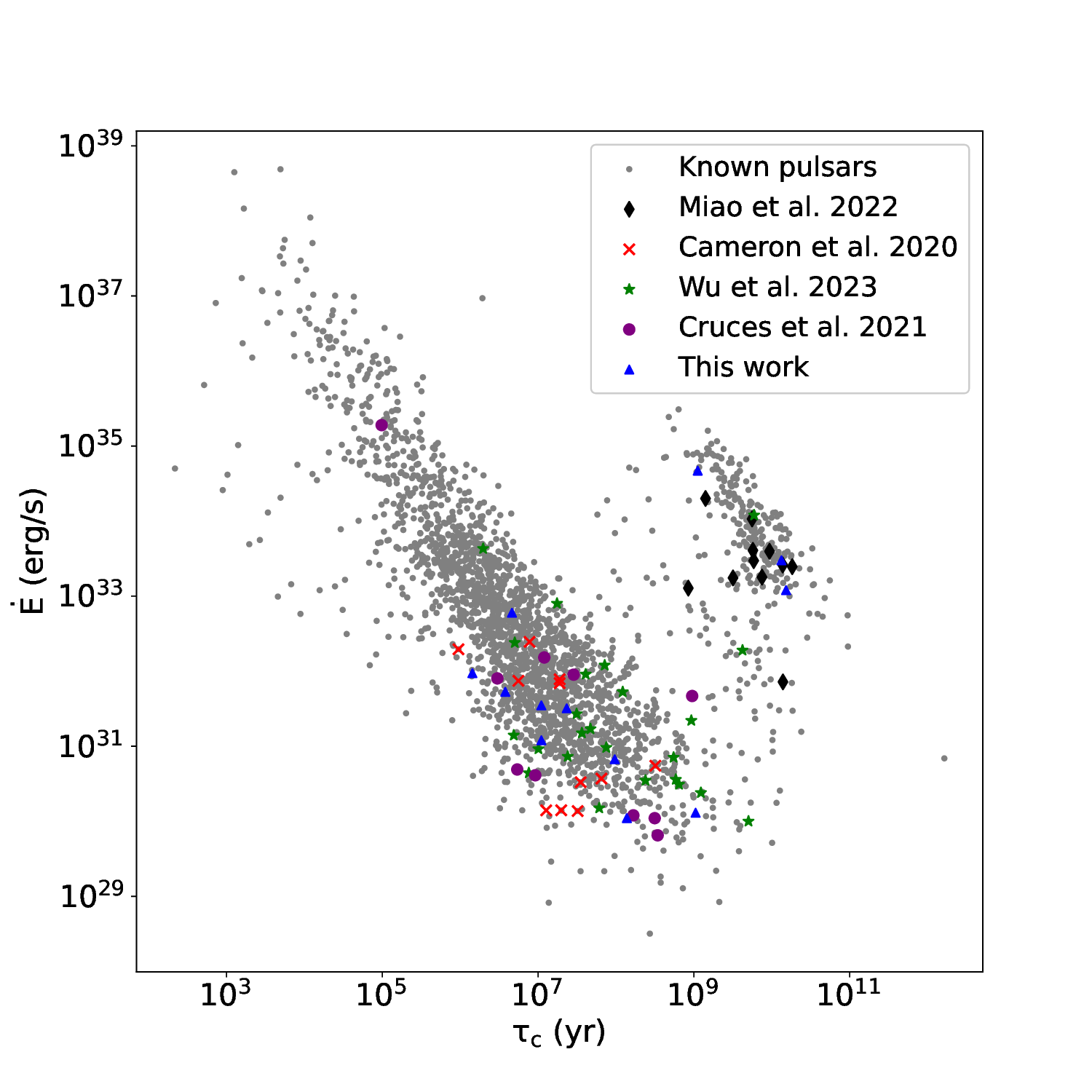}
    \caption{This panel shows the distribution of the spin-down luminosity($\dot{E}$) over characteristic age ($\tau_{c}$). The red stars represent pulsars reported in \citet{2023MNRAS.522.5152W}. The red ``x" stand for CRAFTS pulsars in \citet{2020MNRAS.495.3515C}. The red circles represent pulsars reported in \citet{2021MNRAS.508..300C}. The red rhombuses stand for CRAFTS millisecond pulsars in \citet{2023MNRAS.518.1672M}.
    The blue triangles represent pulsars reported in this work. Other known pulsars are gray dots, and the data is from ATNF (version 1.70).}
    \label{Eyr}
\end{figure}

As shown in panel (a) of Figure~\ref{se}, we present the pulsar $\dot{E}$ and flux ($S_{1400}$) distribution reported for 35 CRAFTS pulsars.
The flux density of 35 pulsars discovered with FAST is significantly lower than that of most known pulsars.
From panel (b) in Figure~\ref{se}, we can see that the new pulsars discovered by FAST have a low flux density within a similar range.
This also indicates that FAST has a very high sensitivity and is very effective for detecting weak sources.
Panel (c) of Figure~\ref{se} shows the dependence of radio luminosity on $\dot{E}$ for 35 pulsars and other known pulsars.
Previous research has shown that the correlation between radio luminosity and $\dot{E}$ is very weak \citep{1993MNRAS.263..403L,2006ARep...50..483M,2014ApJ...784...59S,2020RAA....20..188W}.
From panel (c), it can be observed that these weak sources did not change the weak correlation between L and $\dot{E}$. The panel (d) in Figure~\ref{se} shows the distance distribution of FAST pulsars vs luminosity.
We found that in the same luminosity range, the pulsars discovered by FAST are farther away from us; 
at the same distance, the pulsars found by FAST have lower luminosity. 
In Figure~\ref{Eyr}, we present the spin-down luminosity ($\dot{E}$) vs characteristic age ($\tau_{c}$) distribution of 69 pulsars newly discovered by FAST. It shows that the pulsars newly discovered by FAST are older and $\dot{E}$ is lower than the known pulsars. The pulsar in our work has a lower $\dot{E}$ at the same characteristic age. 
FAST's enhanced sensitivity not only indicates its capability to detect weak sources but also raises the potential to update radiation theory in the future.

\section*{Acknowledgments}

The National Natural Science Foundation of China Grant (No. 12288102, 12041303, 12041304, 12273100, 11988101, T2241020), this is work is supported by the National Key Research and Development Program of China (No. 2022YFC2205203, No. 2022YFC2205202), the  Major Science and Technology Program of Xinjiang Uygur Autonomous Region (No. 2022A03013-3, 2022A03013-4). 
D.L. is a New Cornerstone investigator and is supported by the 2020 project of Xinjiang Uygur autonomous region of China for flexibly fetching in upscale talents. P.W. acknowledges support from the National Natural Science Foundation of China under grant U2031117, the Youth Innovation Promotion Association CAS (id. 2021055), NSFC 11988101 and the Cultivation Project for FAST Scientific Payoff and Research Achievement of CAMS-CAS. WWZ is supported by National SKA Program of China No. 2020SKA0120200, National Nature Science Foundation grant No.11873067. SJD is supported by Guizhou Provincial Science and Technology Foundation (Nos. ZK[2022]304) and  Foundation of Education Bureau of Guizhou Province, China (Grant No. KY (2020) 003). JMY is  Sponsored by Natural Science Foundation of Xinjiiang Uygur Autonomous Region (No. 2022D01D85). 
Tianshan TalentsTraining Program (Scientific and Technological Innovation Team). This work made use of the data from FAST (Five-hundred-meter Aperture Spherical radio Telescope). FAST is a Chinese national mega-science facility, operated by National Astronomical Observatories, Chinese Academy of Sciences. We would like to thank the FAST CRAFTS group and XAO pulsar group for data provide and helpful suggestions that led to significant improvement in our study. R.Y. is supported by the National Key Program for Science and Technology Research and Development No. 2022YFC2205201, the Major Science and Technology Program of Xinjiang Uygur Autonomous Region No. 2022A03013-2, and the open program of the Key Laboratory of Xinjiang Uygur Autonomous Region No. 2020D04049. This research is partly supported by the Operation, Maintenance and Upgrading Fund for Astronomical Telescopes and Facility Instruments, budgeted from the Ministry of Finance of China (MOF) and administrated by the CAS.

\bibliography{sample63}{}
\bibliographystyle{aasjournal}

\end{document}